\documentclass[english]{article}
\usepackage[T1]{fontenc}
\usepackage[latin9]{inputenc}
\usepackage{geometry}
\usepackage{color}
\usepackage{array}
\usepackage{rotfloat}
\usepackage{textcomp}
\usepackage{multirow}
\usepackage{amstext}
\usepackage{graphicx}
\usepackage[authoryear]{natbib}

\makeatletter

\newcommand{\lyxmathsym}[1]{\ifmmode\begingroup\def\b@ld{bold}
  \text{\ifx\math@version\b@ld\bfseries\fi#1}\endgroup\else#1\fi}

\providecommand{\tabularnewline}{\\}

 \date{}

\makeatother

\usepackage{babel}

\begin{document}

\title{Characterization and mapping of surface physical properties of
Mars from CRISM multi-angular data: application to Gusev Crater and
Meridiani Planum}

\author{J. Fernando$^{1,2}$, F. Schmidt$^{1,2}$, C. Pilorget$^{3}$, P. Pinet$^{4,5}$, X. Ceamanos$^{6}$, S. Dout\'e$^{7}$, Y. Daydou$^{4,5}$, F. Costard$^{1,2}$}

\maketitle

\paragraph{\textmd{$^{1}$Univ. Paris-Sud, GEOPS, UMR 8148, Orsay, 91405, France
$^{2}$CNRS, Orsay, 91405, France, jennifer.fernando@u-psud.fr
$^{3}$Division of Geological and Planetary Sciences, Caltech, Pasadena, California, USA
$^{4}$Univ. de Toulouse, UPS-OMP, Toulouse, France
$^{5}$CNRS, IRAP, Toulouse, France
$^{6}$M\'et\'eo France/CNRS, CNRM/GAME, Toulouse, France, $^{7}$Univ. Joseph Fourier/CNRS, IPAG, Grenoble, France}}


\begin{abstract}

The analysis of particle physical properties (grain size,
shape and internal structure) and its organization (surface porosity, roughness, 
and grain size distribution) provides information about the geological processes which
formed and modified planetary surfaces. CRISM (Compact Reconnaissance 
Imaging Spectrometer for Mars) multi-angular observations
(varied emission angles) allow for the characterization of the surface scattering
behavior, which depends on the composition and on the material physical
properties such as the grain size, shape, internal structure, and
the surface roughness. After an atmospheric correction
(compensating mineral aerosol effects) by the Multi-angle Approach
for Retrieval of the Surface Reflectance from CRISM Observations (MARS-ReCO),
the surface reflectances at different geometries were analyzed by inverting
the Hapke photometric model depending on six parameters: single
scattering albedo, 2 phase function terms, macroscopic roughness and
2 opposition effects terms. In this work, surface
photometric maps are created to observe the spatial variations of
surface scattering properties as a function of geological units. Information
regarding the single scattering albedo, the particle phase function and the
macroscopic roughness are provided at the CRISM spatial resolution
(200m/pixel). This article describes the application of this methodology to
the data covering the Mars Exploration Rover (MER) landing sites located at Gusev Crater
and Meridiani Planum where orbital and in situ observations are available.
Complementary orbital observations (e.g., CRISM spectra, THermal EMission
Imaging System (THEMIS), High Resolution Imaging Science Experiment,
(HiRISE) images) were used for interpreting the modeled Hapke
photometric parameters in terms of physical properties which can be used to 
constrain the geological processes. Available MER in situ observations
were used as ground truth to validate the interpretations of the Hapke parameters. Varied scattering
properties were observed within a CRISM observation (5x10km) suggesting
that the surfaces are controlled by local geological processes (e.g.,
volcanic resurfacing, aeolian and impact processes) rather than regional
or global processes. Results are consistent with the in situ observations, thus
validating the approach and the use of photometry for the characterization
of Martian surface physical properties. 

\end{abstract}

\paragraph{\textmd{Citation: Fernando, J. et al. (2015), Characterization and mapping of surface physical properties of Mars from CRISM multi-angular data: application to Gusev Crater and Meridiani Planum, Icarus}}


\section{Introduction}

Planetary surfaces have been modified by a variety of geological and climatic processes, including external (e.g., space weathering), internal (e.g., volcanism) and interactions
between surface and atmosphere (e.g., aqueous and wind erosion, alteration,
transportation and deposition, CO$_{2}$ and H$_{2}$O cycles). These processes are recorded
in the surface mineralogy and the surface texture. The mineral assemblage
gives information about the chemical environment (e.g., crystallization
from an igneous magma, deposition from aqueous agents, recrystallization
due to metamorphism, crystallization during diagenesis process of
sediments, oxidation and weathering of rocks due to interactions with
the atmosphere). Analyses of particle physical properties (grain size, shape and 
internal structure) to its organization (surface porosity, roughness and grain size distribution) 
provide information regarding surface formation and evolution (e.g., the grain size discriminates 
volcanic and plutonic materials) including transportation (e.g., the grain shape reflects the history of
the transportation and/or the capability of the erosion
agent (liquid or wind) to erode the grains), deposition and weathering.

Measurement of the bidirectional reflectance
in visible and near-infrared ranges at different geometries (incidence,
emission and azimuthal angles) provides information about the scattering
behavior. This scattering depends on the composition and on the
material physical properties: from the particle scale such as its
size, shape, internal structure to its organization within a pixel
such as the surface roughness and the grain size distribution. This technique has been used for the
analysis of laboratory measurements \citep[e.g., ][]{mcGuire1995,Piatek2004,cord2005,shepard2007,Pommerol2013,souchon2011,Johnson2013},
and for investigations conducted for Mars using in situ \citep[e.g., ][]{johnson1999,Johnson2006a,Johnson2006b,Johnson2014}
and orbital measurements \citep[e.g., ][]{erard1994,Jehl2008,Fernando2013,Shaw2013}.
In situ measurements provide constraints on the scattering behavior
from specific classes of rocks and soils at 10's of microns to 10's of meters scales and are limited to
the rover path, whereas orbital measurements provide that information
from more extended areas integrating rocks and soils (hectometer to
kilometer spatial scales) and can be obtained everywhere on the planet.
The High Resolution Stereo Camera (HRSC) on-board Mars Express (MEx)
has acquired multi-angular (up to five geometries per orbit) data
of the Martian surface. \citet{Jehl2008} determined the regional
variations of the photometric properties at the kilometer spatial
scale, across Gusev Crater and the south flank of Apollinaris Patera. They used 
several HRSC observations to enrich the diversity of available
geometries (phase angle from 20 to 90$^\circ$). Their photometric study was carried out without any atmospheric
correction but ensuring that the atmospheric contribution was limited
by selecting HRSC observations with mineral aerosol optical thickness
lower than 0.9. Since 2006, the spectro-imager Compact Reconnaissance
Imaging Spectrometer for Mars (CRISM) onboard Mars Reconnaissance
Orbiter (MRO) provides multi-angular hyperspectral images (up to eleven
geometries) giving access to the surface material scattering behavior at the hectometer spatial scale. Recently, \citet{Fernando2013} proposed an approach
to estimate the surface photometric parameters by inverting the Hapke photometric model \citep{Hapke1993,Hapke2012a}, including a robust
and fast atmospheric correction, called the Multi-angle
Approach for Retrieval of the Surface Reflectance from CRISM Observations
(MARS-ReCO) \citep{Ceamanos2012}. The estimated
photometric parameters provided information about the single scattering
albedo and the main direction of the surface scattering (e.g., forward,
backward) directly related to the physical state
of the surface. Moreover, quantitative information regarding the surface
roughness (roughness at microscales scales) could be provided.
In parallel, \citet{Shaw2013} derived maps of the single scattering
albedo and the scattering asymmetry parameter at MER-Opportunity
landing site at Meridiani Planum, focused around the Victoria Crater,
using CRISM multi-angular observations and a coupled surface and atmospheric model.  

In this paper, we apply the methodology developed by \citet{Fernando2013} 
to study the spatial variations of the scattering properties, as a function of 
mapped geological units, in terms of surface photometric parameters. 
We try to link the surface photometric
parameters in terms of physical properties, by coupling with the orbital
observations (e.g., CRISM/MRO spectral criteria, THermal EMission
Imaging System (THEMIS) / Mars Odyssey thermal inertia values, High
Resolution Imaging Science Experiment, (HiRISE)/MRO images and digital
terrain model (DTM)), to provide information about the geological context.
This work focuses on the Mars Exploration Rover (MER) landing
sites, located at Gusev Crater and Meridiani Planum, where orbital and
in situ observations are available. This application is an example
of what the methodology and the mapping can provide as information
about the surface material physical properties. The photometric
results obtained from CRISM data are compared to the MER in situ observations
(e.g., Panoramic camera (Pancam), Microscopic Imager (MI), Navigational
camera (Navcam)), used as ground truth to validate the photometric
interpretations.

The article is organized as follows: (i) the methodology to obtain the 
surface photometric parameters maps from the Hapke model is described
in Section \ref{sec:Methodology} and (ii) the analyses of these maps
are presented and discussed by associating them with complementary
orbital and in situ observations and with laboratory studies in Section
\ref{sec:Results}. Then, we present (iii) the comparisons of the
CRISM photometric results with a numerical model in Section 4 
and (iv) the relations to the geological processes in Section \ref{sec:Geological processes}. 


\section{Data sets and methods \label{sec:Methodology}}

\subsection{CRISM targeted observations}

The CRISM instrument on-board MRO is a visible and
infrared hyperspectral imager (from 362 to 3920 nm at 6.55 nm/channel).
The appropriate mode to estimate the surface spectro-photometric properties
is the targeted mode. We decided to used the Full Resolution Targeted observations
(FRT) which provide the highest spatial resolution (15-20m/pixel), and thus potentially less areal mixtures compared to lower resolution images. 
However, the Half spatial Resolution Long (HRL) and the Half spatial Resolution Short (HRS) observations could also be used by the methodology presented below. A targeted observation is a sequence of 11 hyperspectral images of
a target area, acquired at different emission angles due to the
rotation of the detector at $\pm$70$^\circ$. The solar incidence
is almost constant during the spacecraft flyby over the target. The
typical FRT sequence is composed of a central image ($\sim$10x10
km) at high spatial sampling (15-20m/pixel) and 10 off-nadir images
with a x10 binning (150-200m/pixel), taken before and after the central
image, involving two azimuthal modes \citep{murchie2007}.

\subsection{Methodology \label{sub:Hapke parameter}}

\citet{Fernando2013} presented a methodology for the estimation of the 
Hapke photometric parameters from the CRISM multi-angular observations, summarized below. 
\citet{Fernando2013} also presented a cross-validation of the methodology by 
comparing the CRISM estimates to the in situ photometric results at the MER landing sites. 
Compared to the previous works, the  main goal in this work is to include the mapping of 
the surface photometric parameters in order to study the spatial variations of the scattering 
properties, as a function of mapped geological units, providing extended information over a wider area. 
This work also focuses on the physical 
interpretation of the modeled Hapke photometric parameters. The mapping induces same 
modifications in the previous methodology described in this subsection.

\subsubsection{Integration of multi-angular images}

The eleven hyperspectral images of a FRT observation
(reflectance and ancillary cubes) are spatially rearranged into a
new data set. We refer to this as a spectro-photometric curve (SPC) cube \citep{Ceamanos2012}.
In this work, the integration of the multi-angle images has
been improved (i) by taking into account the geometric deformations (especially at oblique views) caused by the instrument rotation, and (i) by controlling the association between two pixels of two different images of the sequence. Both are not taken into account in the previous method \citep{Ceamanos2012}. 
First, the central image, taken at 15-20m/pixel, is binned
x10 to have the same spatial resolution as the 10 off-nadir images
(150-200m/pixel). Second, all images are spatially combined to build
the spectro-photometric curve of each spatial pixel, by spatially associating
each pixel of the central image (chosen as reference) with pixels
from the 10 off-nadir images. We calculate the spatial nearest pixel
by using the spatial coordinates of each image of the FRT observation,
and by taking into account the geometric deformation due to the detector
gimbal (spatial resolution/cosine(emergence angle)). The association
is performed when the overlapping between two pixels of two different
images (central image and an off-nadir image) is at least 10\%.

\subsubsection{Correction for mineral aerosol contributions and estimation of the
surface bidirectional reflectance}

\citet{Ceamanos2012} proposed a technique, referred
to as MARS-ReCO, to correct for mineral aerosols
exclusively in order to estimate the surface bidirectional reflectance.
For this work, the surface bidirectional reflectance is estimated
at 750 nm where the contribution from gas is negligible. The optical
properties of mineral aerosol (particle size distribution and refractive
index) and the mineral and water ice aerosol optical thickness ($AOT_{mineral}$
and $AOT_{water}$) of each observation are provided by Michael Wolff
(\citet{wolff2009}, personal communication, 2011). The $AOT_{mineral}$
 for each CRISM acquisition is used as an input of MARS-ReCO. MARS-ReCO
is able to propagate the uncertainties on the surface bidirectional
reflectance from the top-of-atmosphere (TOA) measurements. The different
criteria for the use of the MARS-ReCO approach are presented in Sub-section
\ref{sub:Criteria for the CRISM selection}.

\subsubsection{Estimation of Hapke photometric parameters}

To analyze the surface bidirectional reflectance,
the \citet{Hapke1993,Hapke2012} model is used. This model depends
on the geometric angles (incidence, emergence and phase angles) and
on six parameters: the single scattering albedo, the particle phase
function, the surface macroscopic roughness and the opposition effect
parameters (Table \ref{tab:Hapke parameter}). It is important to
mention that CRISM is not observing Mars with phase
angles less than 20$^\circ$. Consequently, the opposition effect
parameters (magnitude $B_{0}$ and  angular width $h$ of the opposition effect 
(Table \ref{tab:Hapke parameter})), are underconstrained. However, as explained by 
\citet{Fernando2013}, neglecting both parameters should not influence the retrieval of the 
other parameters. However, following \citet{souchon2011}, the model can still be 
profitably inverted by keeping parameters $B_{0}$ and $h$. We tested first 
by inverting the parameters $B_{0}$ and $h$ to systematically control if they are 
constrained or not by the CRISM data set using the non-uniformity criterion ($k$) 
described in subsection 2.4. Second we tested the inversion by setting $B_{0}$ 
and $h$ to zero, and no change was observed on the determination of the other parameters 
($\omega$, $\bar{\theta}$, $b$, and $c$). 

The photometric parameters are estimated using a Bayesian inversion framework, adapted 
for the inversion of the non-linear Hapke model
\citep{Fernando2013,Fernando2013b}. This technique is based on the
concept of the state of information, characterized by a probability
density function (PDF) \citep{tarantola1982}. The prior information
about model parameters (no information, defined as uniform PDF) combined
with prior information about observations (a gaussian PDF) are fused
to infer the solution by using Bayes' theory. The final state
of information (posterior PDF of each parameter) is numerically sampled,
using a Monte Carlo Markov Chain \citep{Mosegaard1995}, to calculate the posterior 
PDF and the mean and the standard deviation of each Hapke photometric parameter 
(see \citet{Fernando2013,Fernando2013b} for more details). The advantages of this inversion procedure 
are that the the data uncertainties are taken into account in the model parameter 
estimations and all parameter assemblages are tested.

The single scattering albedo ($\omega$) is related to the composition, the particle 
size and the microstructure (e.g., crystals, fractures, pores) 
\citep{Hapke1993,Hapke2012}. The phase function parameters ($b$ and $c$) are dependent 
and are related to the particle shape, composition and internal structure 
\citep[e.g., ][]{mcGuire1995,souchon2011}. The macroscopic roughness ($\bar{\theta}$) 
parameter is initially defined as the integral surface roughness at the sensor subpixel 
scales \citep{Hapke1993,Hapke2012}. However, several authors demonstrated that the 
surface macroscopic roughness parameter is more sensitive to the microscale (from the 
particle size to a few millimeters) and thus to the grain organization 
\citep{Shepard1998,Helfenstein1999,cord2003,shkuratov2005,shepard2007}. 
In addition to the photometry technique, several techniques and approaches have 
been developed to estimate the Martian surface roughness, specially to have the access 
to the sub-meter scale morphology including, the laser altimeter measurements 
\citep[e.g., ][]{Neumann2003}, the radar measurements \citep[e.g., ][]{Campbell2001} and 
the surface temperature measurements \citep[e.g., ][]{Bandfield2008}. However, all these 
approaches did not provide the same scale of surface roughness. The surface roughness 
is characterized at scales larger than about 300 m from laser altimeter measurements, 
like Mars Orbiter Laser Altimeter (MOLA) data, at centimeter to meters scales from radar observations, and at 0.01 m  
(for low thermal inertia to 0.1 m for moderate thermal inertia) from thermal infrared 
observations. Consequently, a direct comparison is not possible due to the different 
roughness scales but all are complementary. 


\begin{table}
\begin{centering}
{\scriptsize }%
\begin{tabular}{>{\centering}p{2cm}>{\centering}m{4cm}>{\centering}m{4cm}>{\centering}m{4cm}}
\hline 
{\scriptsize Symbole} & {\scriptsize Definition} & {\scriptsize Physical significance} & {\scriptsize Geological significance}\tabularnewline
\hline 
{\scriptsize $\omega$} & {\scriptsize Single scattering albedo} & {\scriptsize ratio of scattered light at the particle
scale to extincted light} & {\scriptsize particle composition and size and microstructure}\tabularnewline
{\scriptsize $b$} & {\scriptsize Asymmetry parameter$^{a}$} & {\scriptsize anisotropy of the scattering ($b<0.5$:
broad / $b\geq0.5$: narrow scattering lobe)} & {\scriptsize particle shape, composition and internal structure}\tabularnewline
{\scriptsize $c$} & {\scriptsize Backscattering fraction$^{a}$ } & {\scriptsize main scattering direction ($c<0.5$:
forward / $c\geq0.5$: backward scattering) } & {\scriptsize particle shape, composition and internal structure}\tabularnewline
{\scriptsize $B_{0}$} & {\scriptsize Magnitude of the opposition effect} & {\scriptsize magnitude of the opposition effect peak ($g<$5$^\circ$)} & {\scriptsize particle transparency ($B_{0}=0$: transparent
particle, $B_{0}=1$: opaque particle) }\tabularnewline
{\scriptsize $h$} & {\scriptsize Angular width of the opposition effect} & {\scriptsize angular half width of the opposition effect peak ($g<$5$^\circ$)} & {\scriptsize surface porosity, particle size ($h=0$:
high porosity, $h=1$: low porosity and/or more uniform grain size distribution)}\tabularnewline
{\scriptsize $\bar{\theta}$} & {\scriptsize Macroscopic roughness} & {\scriptsize mean slope angle within a pixel expressed
in degree} & {\scriptsize roughness at microscale (from the particle to a few mm)}\tabularnewline
\hline 
\end{tabular}
\par\end{centering}{\scriptsize \par}

{\scriptsize $^{a}$We assume a two-term Henyey-Greenstein phase function
(HG2) }{\scriptsize \par}

{\scriptsize $g$: phase angle}{\scriptsize \par}

\caption{Summary of the Hapke photometric parameters and their physical and
geological significance. All parameters are depending
of the wavelength, set to 750 nm in this work. Note that the
parameters $\omega$, $b$, $c$, $B_{0}$ and $h$ vary from 0 to
1 and the parameter $\bar{\theta}$ varies from 0 to 45$^\circ$.
\label{tab:Hapke parameter} }
\end{table}


\subsection{Criteria for the selection of CRISM observations \label{sub:Criteria for the CRISM selection} }

Since September, 2010, the inbound segment in CRISM targeted mode
is absent due to problems of the gimbal instrument \citep{Murchie2012}.
In order to have the maximum number of geometries, only CRISM observations
acquired before this date are selected. 

The MARS-ReCO procedure is suitable for any CRISM observation within
the following constraints described below.

(1) The accuracy of the determination of the surface reflectance highly depends on
the combination of a moderate mineral aerosol opacity ($AOT_{mineral}$
$\leq$ 2, \citep{Ceamanos2012}) and a low water ice aerosol opacity,
as the latter is not corrected by MARS-ReCO ($AOT_{water}$ $\leq$
0.2, water ice content). 

2) The accuracy of the determination of the surface reflectance highly depends on the geometries. 
The MARS-ReCO procedure is suitable for any CRISM multi-angular
observation within these geometrical constraints: incidence angle $\theta_{0}$
$<$ 60$^\circ$, phase angle range $\Delta g=g_{max}-g_{min}>$40$^\circ$,
outside the plane perpendicular to the principal plane ($\varphi_{inbound}=\varphi_{outbound}\sim$ 90$^\circ$)
\citep{Ceamanos2012}. The local topography makes the photometric
study more challenging when it is poorly known, because it controls
to a large extent, the incidence, emergence, and azimuth local angles.
Besides, in the case of an oblique illumination (i.e., up to 70$^\circ$),
shadows decrease the signal/noise ratio. Consequently, regions with steep surfaces which 
can be observed within a CRISM central image (10kmx10km) but which may create high shadows, 
are excluded for this work (e.g., the crater central peaks, slopes and rims, the channel 
slopes, the steep hills), to minimize the errors. The regions with high relief are 
evaluated from the high resolution HiRISE images or from the HiRISE digital terrain 
models and their photometric results are masked in the final maps. 

The success of the MARS-ReCO procedure is evaluated by the percentage
of corrected pixels. Observations with more than 50\% of failed pixels after the 
MARS-ReCO procedure are rejected (a failed pixel means that the MARS-ReCO did not 
correct its photometric curve). The failure can be explained by the fact that one or 
several geometric and/or atmospheric conditions, detailed previously, is/are not respected.

\subsection{Evaluation on the accuracy of the surface photometric parameters
estimates}

After the MARS-ReCO correction, we can evaluate if the surface photometric
curve (especially the diversity and number of available geometries)
provides enough information to estimate accurate Hapke photometric
parameters. For that purpose, we analyze the shape of the posterior PDF. The
criteria are summarized below. 

(1) Non-uniformity criterion ($k$): A solution exists if the posterior
PDF differs from the prior information (a uniform distribution). A
statistical test is performed, leading to a non-uniformity criterion
$k$ \citep{Fernando2013}. For $k\geq0.5$, the posterior PDF is
considered to be a non-uniform PDF, meaning that a solution exists.

(2) Bimodality of the single scattering albedo PDF criterion: The
presence of two possible solutions (e.g., bimodal distribution) for the $\omega$ parameter is
the consequence of the limitation of geometric diversity in the CRISM
photometric curve to constrain the $\omega$ parameter \citep{Fernando2013,Fernando2013b},
which is usually the best-constrained parameter in photometric modeling.
Synthetic tests showed that if a pixel has a bimodal distribution for the  $\omega$ parameter, generally one or several parameters has/have no solution or if a solution exists, this solution is underconstrained (i.e., high standard deviation) \citep{Fernando2013,Fernando2013b}. All pixels having a bimodal distribution for the $\omega$ parameter are not considered and the photometric parameter set are not used and not mapped. 

(3) Standard deviation ($\sigma$) criterion: This is used to characterize
the dispersion around the mean value, giving constraints on the
accuracy of the solution. A solution is considered as well-constrained
when $\sigma_{\omega}\leq0.10$ for the $\omega$ parameter (which
is the best constrained parameter in modeling), $\sigma_{b,c}\leq0.20$
for the $b$ and $c$ parameters and $\sigma_{\bar{\theta}}\leq$5$^\circ$
for the $\bar{\theta}$ parameter ($b$, $c$ and $\bar{\theta}$ are the less well-constrained
parameters due the limited range of CRISM geometries). 

The quality of the eleven multi-angular images comprising the region under study is 
crucial. Indeed, the ten off-nadir image footprints must ideally overlap the central
image. However, due the surface topography uncertainty, the spacecraft
position and attitude uncertainties, and the gimbal jitter of the instrument,
the overlap among all images is not always ideal. Consequently, a worst
quality of the image overlap reduces the number of available
angular configurations and the phase angle range. Only pixels with
at least seven angular configurations are used for the estimation
of surface photometric parameters. Tests using synthetic data sets
showed that the sampling of photometric curve with less than seven geometries 
(among those available with CRISM observations) is not sufficient to accurately 
constrain the photometric parameters.

If all the above criteria are not respected, that means
that the diversity of geometries and the sampling of the surface bidirectional
reflectance are not enough to constrain all the photometric parameters.
The solution, proposed by \citet{pinet2005,Jehl2008,Fernando2013},
is to combine several observations, acquired under varied illumination
conditions (assuming no surface changes) to improve the diversity
of geometries.

In spaceborne and laboratory photometric data sets, interplay between certain 
parameters can lead to the same photometric curve. For instance, \citet{shepard2007} 
observed that other Hapke parameters (such as the opposition effect parameters and 
the phase function parameters) can mimic the phase angle effects of macroscopic surface 
roughness at small phase angles. To reduce this effect, several authors showed that a 
high diversity of geometries (a broad phase angle range containing low and high phase 
angles) is necessary for constraining as possible all the parameter set 
\citep[e.g., ][]{Helfenstein1999,shepard2007,cord2003}. These degeneracy problem cannot 
be handled with usual inversion method based on criteria minimization, leading to one single solution. 
Therefore we applied a Bayesian inversion approach in order to test all possibilities and to keep all solutions \citep{Fernando2013}. 
Within this framework, the solution is described as a probability density function on the parameter space. 
We demonstrated that phase angles higher than 90 degrees and large phase angle ranges ($>$ 50 degrees) 
are required to limit the degeneracy in order to constrain all Hapke parameters 
\citep{Fernando2013,Fernando2013b}. In this work, all selected CRISM 
observations have an available highest phase angle $>$ 100 degrees, 
a wide range of phase angles (Table \ref{tab:Selected CRISM FRT}) and a number of 
geometries greater than 7, which all minimize the tradeoff between the model parameters. 
Moreover, the Bayesian inversion and the presented criteria enumerated above are 
suitable for evaluating the quality of the parameter estimates compared to the sampling 
and the diversity of geometries of the surface photometric curve.


\begin{table}
\begin{centering}
{\scriptsize }%
\begin{tabular}{c>{\centering}p{4.2cm}cccc}
\hline 
 &  & {\scriptsize MER-Spirit} & \multicolumn{3}{c}{{\scriptsize MER-Opportunity}}\tabularnewline
 & {\scriptsize ID } & {\scriptsize FRT\#C9FB} & {\scriptsize FRT\#B6B5} & {\scriptsize FRT\#334D} & {\scriptsize FRT\#193AB}\tabularnewline
\hline 
\multirow{2}{*}{{\scriptsize Time}} & {\scriptsize Acquisition date} & {\scriptsize 2008-09-21} & {\scriptsize 2008-07-08} & {\scriptsize 2006-11-30} & {\scriptsize 2010-06-09}\tabularnewline
 & {\scriptsize $Ls$(deg.)} & {\scriptsize 130.321} & {\scriptsize 96} & {\scriptsize 142.97} & {\scriptsize 102.16}\tabularnewline
\cline{2-6} 
\multirow{4}{*}{{\scriptsize Geometry}} & {\scriptsize $\theta_{0}$ (deg.)} & {\scriptsize 63} & {\scriptsize 56} & {\scriptsize 55} & {\scriptsize 56}\tabularnewline
 & {\scriptsize $\varphi_{in}-\varphi_{out}$ (deg.)} & {\scriptsize $\simeq$ 55 - 131} & {\scriptsize $\simeq$ 44 - 128} & {\scriptsize $\simeq$ 64 - 117} & {\scriptsize $\simeq$ 50 - 128}\tabularnewline
 & {\scriptsize $g_{min}$-$g_{max}$ (deg.)} & {\scriptsize $\simeq$ 45 - 110} & \textbf{\scriptsize $\simeq$}{\scriptsize{} 40 - 106} & {\scriptsize $\simeq$ 49 - 97} & {\scriptsize $\simeq$ 39 - 105}\tabularnewline
 & {\scriptsize{} $\Delta g$ (deg.)} & {\scriptsize 65} & \multirow{1}{*}{{\scriptsize 66}} & {\scriptsize 48} & {\scriptsize 66}\tabularnewline
\cline{2-6} 
\multirow{2}{*}{{\scriptsize Atmosphere$^{1}$}} & {\scriptsize $AOT_{mineral}$ (900 nm)} & {\scriptsize 0.25$\pm$0.03} & {\scriptsize 0.35$\pm$0.04} & {\scriptsize 0.35$\pm$0.04} & {\scriptsize 0.31$\pm$0.03}\tabularnewline
 & {\scriptsize $AOT_{water}$ (320 nm)} & {\scriptsize 0.07$\pm$0.03} & {\scriptsize 0.14$\pm$0.03} & {\scriptsize 0.12$\pm$0.03} & {\scriptsize 0.13$\pm$0.03}\tabularnewline
\cline{2-6} 
\multirow{2}{*}{{\scriptsize MARS-ReCO}} & {\scriptsize uncorrected pixels (\%)} & {\scriptsize 9} & \multicolumn{2}{c}{{\scriptsize 33}} & {\scriptsize 13}\tabularnewline
 & {\scriptsize nb. of corrected pixels} & {\scriptsize 868} & \multicolumn{2}{c}{{\scriptsize 1703}} & {\scriptsize 1492}\tabularnewline
\cline{2-6} 
\multirow{9}{*}{{\scriptsize Bayesian inversion}} & {\scriptsize nb. of pixels with $k_{b}\geq0.5$ (\%)} & {\scriptsize 100} & \multicolumn{2}{c}{{\scriptsize 99}} & {\scriptsize 100}\tabularnewline
 & {\scriptsize nb. of pixels with $k_{c}\geq0.5$ (\%)} & {\scriptsize 100} & \multicolumn{2}{c}{{\scriptsize 99}} & {\scriptsize 89}\tabularnewline
 & {\scriptsize nb. of pixels with $k_{\bar{\theta}}\geq0.5$ (\%)} & {\scriptsize 100} & \multicolumn{2}{c}{{\scriptsize 100}} & {\scriptsize 100}\tabularnewline
 & {\scriptsize nb. of pixels with $k_{\omega}\geq0.5$ (\%)} & {\scriptsize 100} & \multicolumn{2}{c}{{\scriptsize 100}} & {\scriptsize 100}\tabularnewline
\cline{2-6} 
 & {\scriptsize nb. of pixels with a bimodality PDF for $\omega$ (\%)} & {\scriptsize 25} & \multicolumn{2}{c}{{\scriptsize 11}} & {\scriptsize 17}\tabularnewline
\cline{2-6} 
 & {\scriptsize nb. of pixels with $\sigma_{b}\leq0.20$ (\%)} & {\scriptsize 55} & \multicolumn{2}{c}{{\scriptsize 72}} & {\scriptsize 80}\tabularnewline
 & {\scriptsize nb. of pixels with $\sigma_{c}\leq0.20$ (\%)} & {\scriptsize 72} & \multicolumn{2}{c}{{\scriptsize 51}} & {\scriptsize 36}\tabularnewline
 & {\scriptsize nb. of pixels with $\sigma_{\bar{\theta}}\leq$5$^\circ$
(\%)} & {\scriptsize 78} & \multicolumn{2}{c}{{\scriptsize 43}} & {\scriptsize 42}\tabularnewline
 & {\scriptsize nb. of pixels with $\sigma_{\omega}\leq0.1$ (\%)} & {\scriptsize 100} & \multicolumn{2}{c}{{\scriptsize 100}} & {\scriptsize 97}\tabularnewline
\hline 
\end{tabular}
\par\end{centering}{\scriptsize \par}

{\scriptsize $Ls$: Solar longitude, $\theta_{0}$: incidence angle,
$\varphi_{in}-\varphi_{out}$: CRISM inbound and outbound azimuthal
angles, $g_{min}$-$g_{max}$: minimum and maximum of the phase angles,
$\Delta g$: phase angle range, $AOT_{mineral}$: mineral aerosol
optical thickness, $AOT_{water}$: water ice aerosol optical thickness,
$\omega$: single scattering albedo, $b$: asymmetric parameter, $c$:
backscattering fraction, $\bar{\theta}$: macroscopic roughness, conf.:
angular configurations, $k$: non-uniform criterion, $\sigma$: standard
deviation, PDF: probability density function, $^{1}$from Wolff's
estimates \citep[personal communication]{wolff2009}}{\scriptsize \par}

\caption{Selected CRISM FRT observations at the MER-Spirit's and MER-Opportunity's
landing sites, respectively at Gusev Crater and Meridiani Planum with
information about geometric, atmospheric conditions and statictic results
relative to the MARS-ReCO procedure and to the Bayesian inversion
of the Hapke model. \label{tab:Selected CRISM FRT}}
\end{table}



\section{Results of CRISM photometric maps \label{sec:Results}}

\subsection{MER-Spirit landing site at Gusev Crater \label{sub:MER Spirit}}

\subsubsection{Selection of CRISM observations \label{sub:MER Spirit selection}}

Up to eleven FRT observations are available at the MER-Spirit landing
site from the beginning of CRISM operations to September 2010. Among
all, the observation FRT\#C9FB shows the best satisfactory statistical
results relative to the MARS-ReCO procedure and to the Bayesian inversion
(satisfactory values of the non-uniform criterion ($k$), the standard
deviation criterion ($\sigma$) and the bimodality of the single scattering
albedo PDF criterion)  (Table \ref{tab:Selected CRISM FRT}). The solutions estimated from the observation
FRT\#C9FB are used in this study.

\subsubsection{Geological context and study areas}

Gusev Crater is an impact structure from the Noachian
epoch, approximately 160 km in diameter and centered at 14.5$^\circ$ S/175$^\circ$ E
\citep{Kuzmin2000}. Previous studies suggested past fluvial and lacustrine
activity from the 800-km long canyon Ma'adim Vallis, with sediment
deposition that was at least partly responsible for the formation of the Columbia Hills 
\citep{Kuzmin2000,Cabrol2003}. However, some other authors recognized multiple eruptions 
of fluid basalts \citep{Greeley2005}, 
analogous to mare basalts \citep{Greeley1993}. The region is also
affected by seasonal aeolian processes. Hundreds of dark-toned, small
sub-parallel streaks are observed. The tracks represent the removal
of fine, bright-toned materials (consistent with dust \citep{Martinez-Alonso2005}) from 
the basaltic underlying dark-toned materials by dust devils \citep{Greeley2003} and/or storms. 

\begin{figure}[H]
\begin{centering}
\includegraphics[scale=0.60]{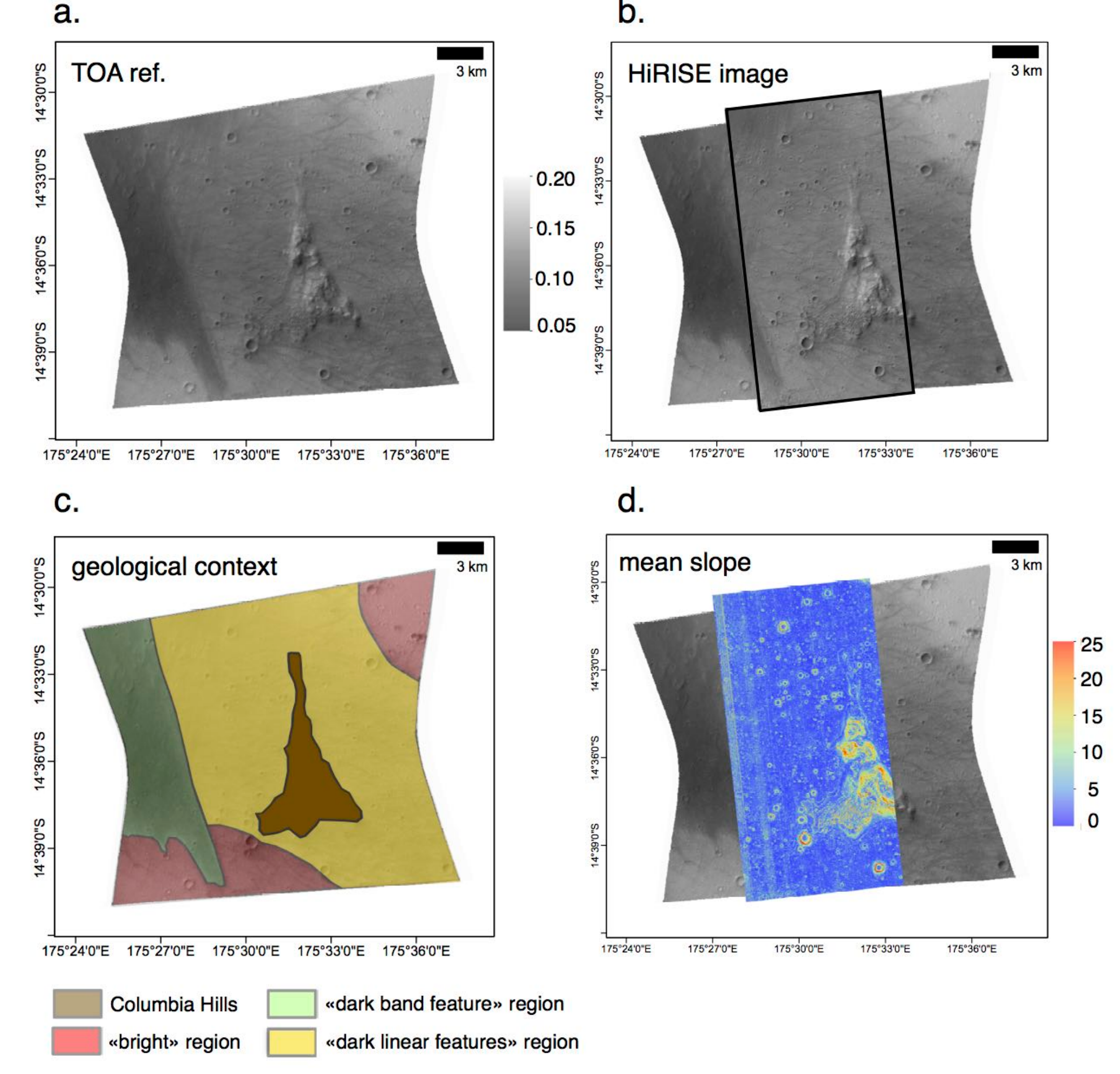}
\par\end{centering}

\caption{Geological context and studied area: a. CRISM top-of-atmosphere (TOA) central images (FRT\#C9FB)
at 20m/pxl at 750 nm, b. CRISM TOA central images overlapped by the
associated HiRISE image (PSP\_010097\_1655\_RED), c. CRISM TOA overlapped by geological context 
map showing the different units and structures, d. CRISM TOA overlapped by the mean slope map at 1 m derived
from HiRISE Digital Terrain Model (DTM) (DTEEC\_001513\_1655\_001777\_1650\_U01). \label{fig: context-map-gusev}}
\end{figure}

The context region of our work is presented in Figure \ref{fig: context-map-gusev}: 
the selected CRISM FRT\#C9FB observation 
is presented in Figure \ref{fig: context-map-gusev}a and the geological 
map, summarizing the different units and structures observed in the 
coupled HiRISE image (Figure \ref{fig: context-map-gusev}b), is illustrated 
in Figure \ref{fig: context-map-gusev}c. The HiRISE DTM image was 
used to calculate the mean slope at the meter per pixel (Figure \ref{fig: context-map-gusev}d). 
Two units are discernible: the hills (Figure \ref{fig: context-map-gusev}c 
in brown color), and the flat plain which is heavily cratered (referred 
to as Gusev cratered plain). Since the hills present 
high local slopes, it would require additional topographic modeling (beyond the scope of 
the current effort) to warrant their inclusion in the models. As such, this unit is masked 
in all subsequent model results. In detail, three units 
are observed: 

(1) a dark-toned large band (NW-SE direction) associated
with a dust storm track, referred to as ``dark band feature''  
(Figure \ref{fig: context-map-gusev}c, in green color),

(2) a region composed of numerous dark sub-parallel
linear features (W-E direction) associated with dust devil tracks,
referred to as ``dark linear features'' region (Figure
\ref{fig: context-map-gusev}c, in yellow color),

(3) a bright-toned region is observed in the NE and SW 
part of the CRISM observations, referred to as ``bright-toned''
region (Figure \ref{fig: context-map-gusev}c, in red
color).

The thermal inertia (TI) provides constraints on
the bulk density, the particle size and the cohesion. From thermal
emission imaging system (THEMIS) onboard Mars Odyssey, the average TI was estimated
as 240 $\pm$ 20 $J.m^{-2}.s^{-0.5}.K^{-1}$ (spatial resolution
of 100m/pixel) by \citet{Milam2003}. The authors interpreted it as surface
of medium-sized grain of sand. This value is associated with the Low Albedo ($LA_{t}$) unit 
\citep{Milam2003}, the region where the selected CRISM observation is located. 

\begin{figure}[H]
\begin{centering}
\includegraphics[scale=0.50]{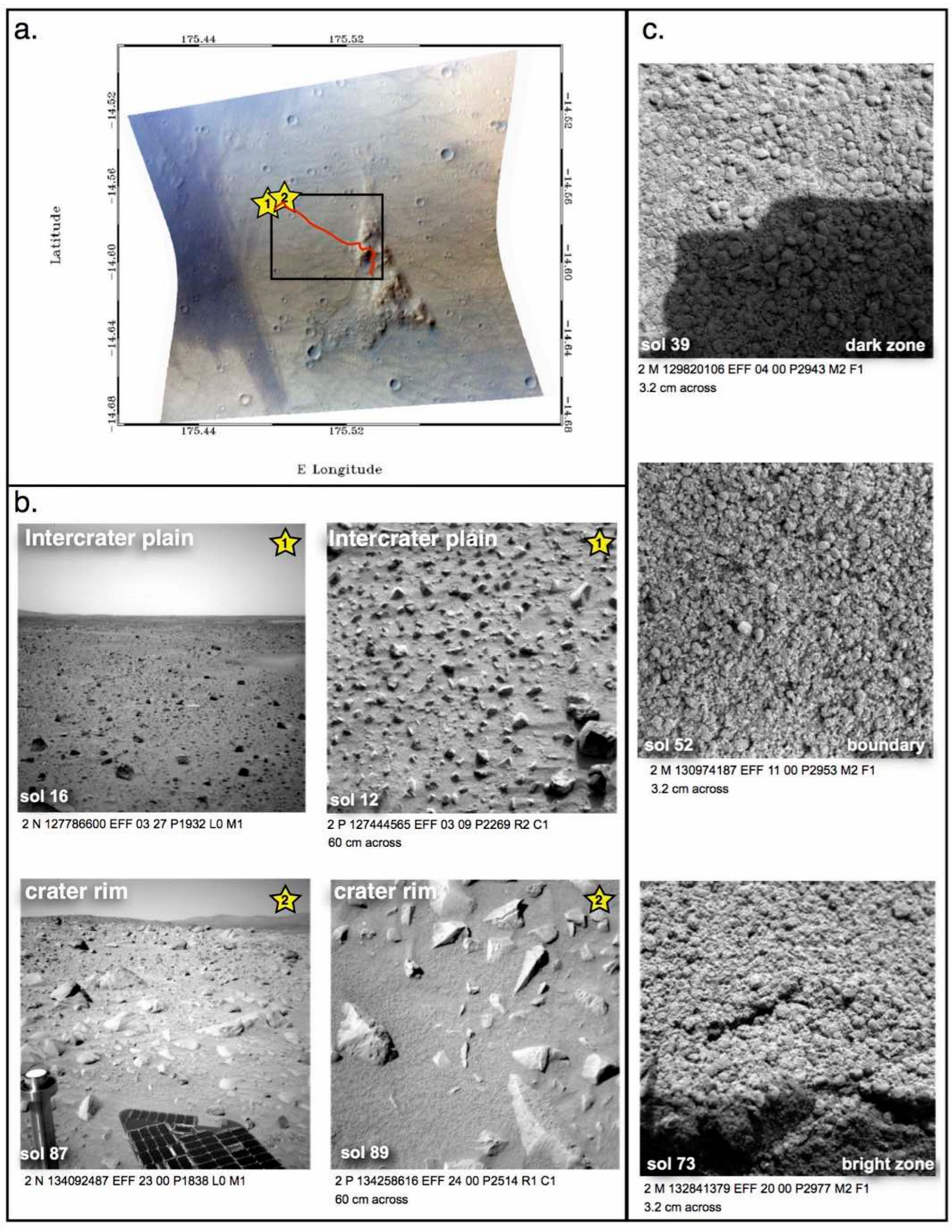}
\par\end{centering}

\caption{In situ observations from navigational camera (Navcam)/MER (N), Panoramic camera (Pancam)/MER (P) and Microscopic Imager (MI)/MER
(M). (a) FRT\#C9FB RGB image at $\sim$20m/pxl (credit: http://crism-map.jhuapl.edu/)
with the rover path (in red color) and the locations of selected in
situ images (location 2 is near Bonneville crater), (b) Navcam and Pancam images in the intercrater region
(top) and near crater rim (bottom) and (c) MI images focused on inside
a dust devil track (top), in the boundary (middle), and outside the
dust devil track (bottom). \label{fig: in-situ-observation-gusev}}
\end{figure}

In situ observations acquired by the MER-Spirit rover provide information
about the surface physical and chemical properties from meter- to millimeter-
scales of main geological units. \citet{arvidson2006a} and \citet{Arvidson2008}
provided an overview of key observations of soils and rocks.  Along the traverse in the Gusev plains 
(Figure \ref{fig: in-situ-observation-gusev}), images
showed that the soil is composed of: (i) basaltic material excavated
from lava flows by local impacts \citep{Greeley2005}, (ii) a layer
of dark coarse sands and granules (0.5 to a few mm in diameter) \citep{herkenhoff2004a,Herkenhoff2006} (Figure
\ref{fig: in-situ-observation-gusev}c, top), interpreted to be rich-olivine basalts 
\citep{bell2004a, Christensen2004b,McSween.2004}, (iii) subangular
lithic fragments, interpreted to be ejecta deposits, which are associated with 
the numerous craters, composed of clasts and rocks, and (iv) all coated
with bright fine-grained materials, inferred to be dust with particles from silt ($<4$ $\mu m$ in
diameter) \citep{lemmon2004} to fine sand (mostly less than 150 $\mu m$),
forming dust aggregates, \citep{herkenhoff2004a,Sullivan2008,Vaughan2010} 
(Figure \ref{fig: in-situ-observation-gusev}c, bottom). 


\subsubsection{Surface material photometric parameters \label{sub:MER spirit parameters}}

\subsubsection*{3.1.3.1. The single scattering albedo parameter \label{sub:MER Spirit w} }

\begin{figure}[H]
\begin{centering}
\includegraphics[scale=0.45]{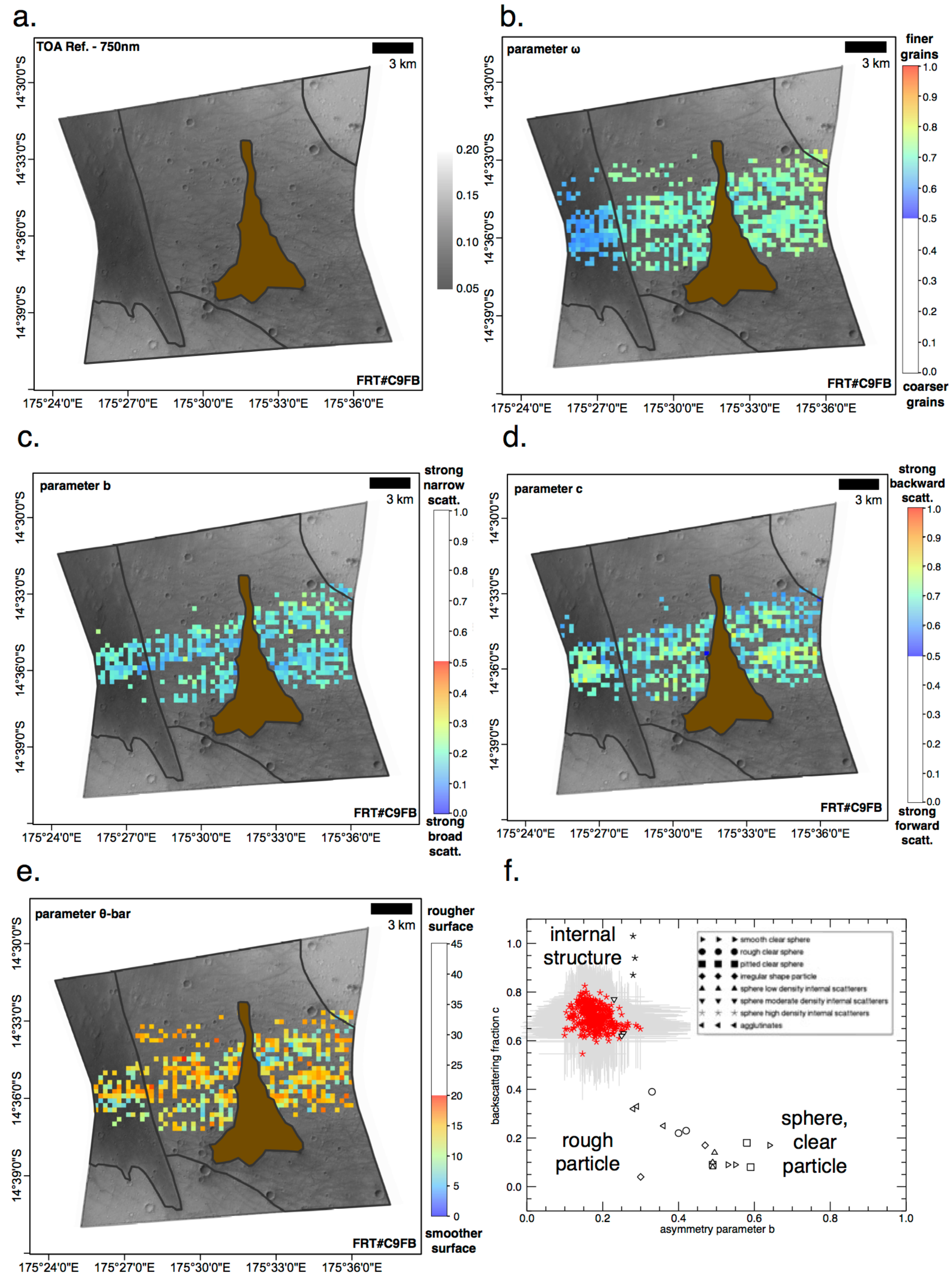}
\par\end{centering}

\caption{CRISM mapping (FRT\#C9FB). a. TOA reflectance map in I/F unit at 20m/pixel
at 750 nm. b. parameter $\omega$ map at 200m/pixel ($\sigma_{\omega}\leq0.10$).
c. parameter $b$ map at 200m/pixel ($\sigma_{b}\leq0.20$). d. parameter
$c$ map at 200m/pixel ($\sigma_{c}\leq0.20$). e. parameter $\bar{\theta}$
map at 200m/pixel ($\sigma_{\bar{\theta}}\leq$5$^\circ$).
The colored pixels correspond to the value of the mean PDF. Only the middle part of the central image is covered with
all additional geometric images (up to 11 images) that it is why the
photometric results are obtained in this area. f. graph of the asymmetry
parameter ($b$) versus backscattering fraction ($c$) estimated from
FRT\#C9FB plotted with experimental values on artificial particles
from \citet{mcGuire1995}. Note that the results in the hills unit are masket out due to high local topography. \label{fig: orbital-parameters-gusev}}
\end{figure}

Figure \ref{fig: orbital-parameters-gusev}b represents the map of
the single scattering albedo parameter ($\omega$) values. Spatial
variations of the $\omega$ parameter are observed at $\sim$200 meters
CRISM scale: (i) the ``dark band feature'' region (Figure
\ref{fig: context-map-gusev}c, in green color) is correlated to the
lowest $\omega$ values ($\omega<0.60$, $\sigma\leq0.10$), (ii)
the ``dark linear features'' region (Figure \ref{fig: context-map-gusev}c,
in yellow color) is correlated to intermediate $\omega$ values ($0.65<\omega<0.75$,
$\sigma\leq0.10$), (iii) the ``bright-toned' region (Figure
\ref{fig: context-map-gusev}c, in red color) is associated with the
highest $\omega$ values ($\omega>0.75$, $\sigma\leq0.10$). 

\begin{figure}[H]
\begin{centering}
\includegraphics[scale=0.45]{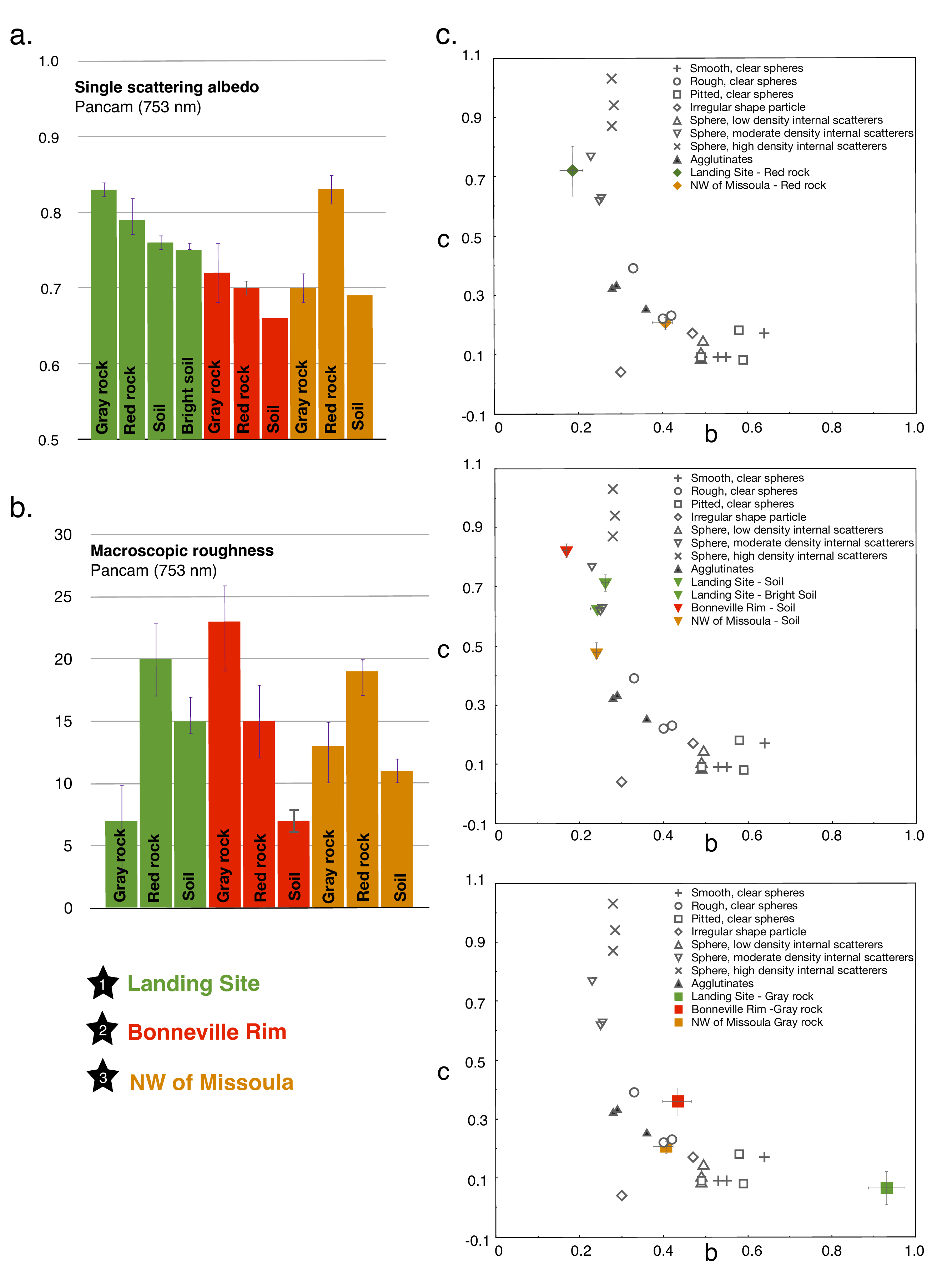}
\par\end{centering}

\caption{In situ photometric results from Pancam/MER: Mean and uncertainties
of a. the single scattering albedo ($\omega$), b. the macroscopic
roughness ($\bar{\theta}$), c. the particle phase function parameters
($b$ and $c$) overplot to the experimental $b$ and $c$ values
pertaining to artificial particles measured by \citet{mcGuire1995}.
All photometric parameters are estimated at 753 nm (except for the parameter $c$ of Red Rock unit for landing site which corresponds to the 754 nm model results) from Pancam onboard Spirit for different
geological units at Landing Site (Sol 013), Bonneville rim (Sol 087-088),
and NW of Missoula (Sol 102-103) \citep{Johnson2006a}. \label{fig: in-situ-parameters-gusev}}
\end{figure}

The $\omega$ values estimated from CRISM ($\omega_{CRISM}$ $\simeq0.65-0.75$) are compared to those estimated
from Pancam measurements at 753 nm, located in our ``dark-linear feature''
geological unit, for different geological units (rocks and soils) \citep{Johnson2006a} 
(Figure \ref{fig: in-situ-parameters-gusev}a). 
The $\omega$ values are consistent with those obtained at Bonneville Rim site ($\omega$ $_{Pancam}\simeq0.66$), Landing Site and NW of Missoula areas ($\omega$ $_{Pancam}\simeq0.75$). 

\begin{figure}[H]
\begin{centering}
\includegraphics[scale=0.40]{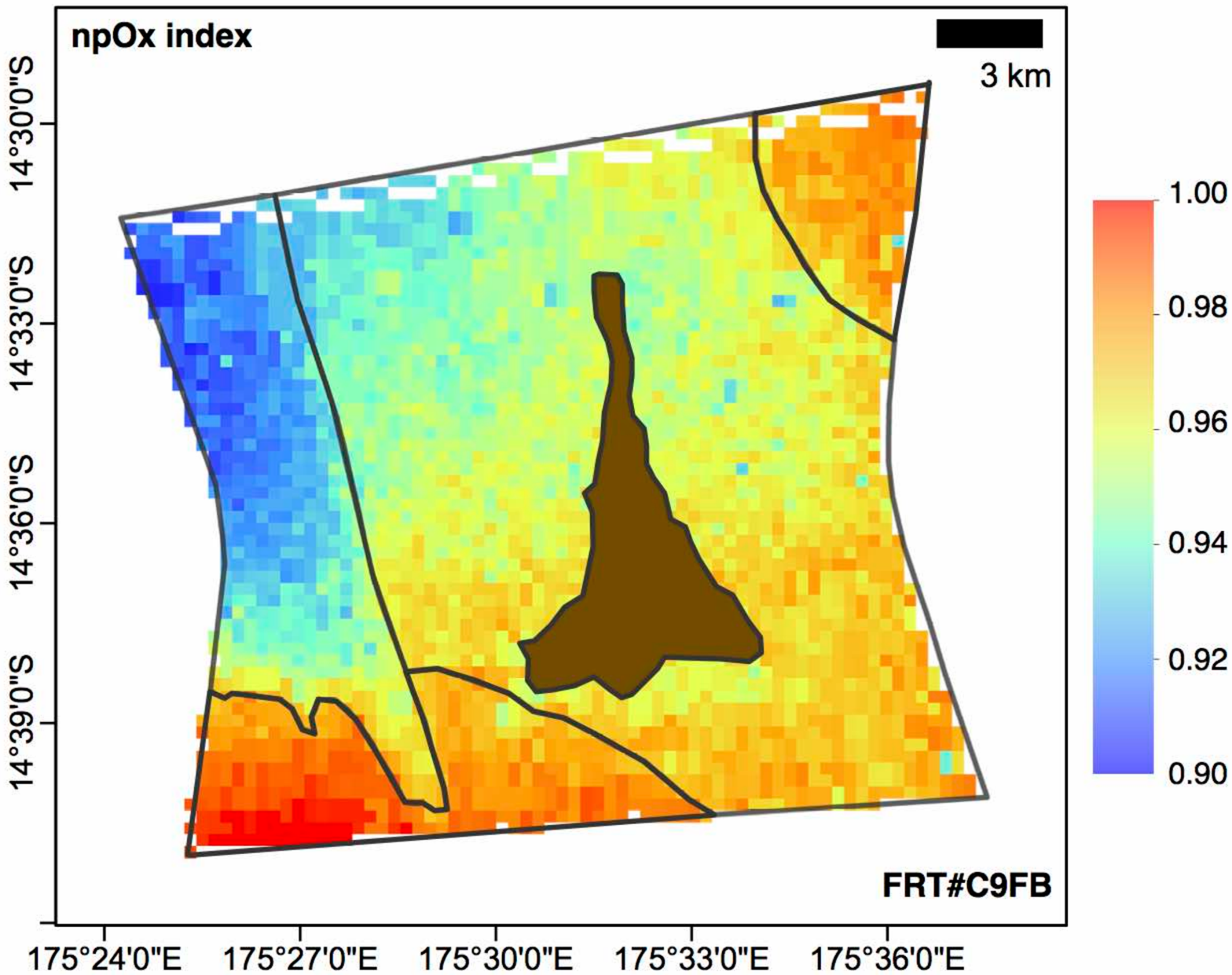}
\par\end{centering}

\caption{Composition map of nanophase ferric oxide spectral index (dust index)
from FRT\#C9FB: low or no dust areas have dust index less than 1 (blue
tones) whereas dusty region have dust index greater or equal to 1
(red tones). The spatial resolution is 200m/pixel. \label{fig: orbital-parameters-gusev-composition}}
\end{figure}

Orbital observations showed that the Gusev Crater
plain is composed of basalt, detected in the thermal infrared range
by the Thermal Emission Spectromer (TES) instrument \citep{Milam2003}. Moreover, nanophase ferric-oxide-rich
(npOx) bright dust is detected, which partially covers the basaltic substrate
detected by the Observatoire pour la Min\'eralogie, l'Eau, les Glaces
et l'Activit\'e (OMEGA) instrument over
the MER-Spirit landing site \citep{Lichtenberg2007}. Those results were confirmed by
the in situ observations. To assess the distribution of surface dust, the nanophase
ferric oxide (npOx) spectral index was calculated from the CRISM data set and mapped 
(Figure \ref{fig: orbital-parameters-gusev-composition}). Based
on the presence of a broad near-infrared absorption between 0.75 and
1.0 $\mu m$ due to the Fe$^{3+}$ electronic transitions, the spectral
ratio of the reflectance at wavelengths 1 $\mu m$ and 0.8 $\mu m$
is calculated (so called 1/0.8 $\mu m$ slope) to estimate the nature
of ferric oxide \citep{Poulet2007}. Values close
to 1 or greater are related to the high contamination of dust which
masks the underlying materials. However, the npOx index is not corrected for mineral aerosol 
contribution, and is therefore sensitive to the
mineral aerosol content in the atmosphere. The $AOT_{mineral}$
estimated for the FRT\#C9FB is relatively low (0.25$\pm$0.03, 
Table \ref{tab:Selected CRISM FRT}), which suggests that the highly structured spatial 
distribution of dust index in Figure \ref{fig: orbital-parameters-gusev-composition}, 
dominantly shows surface signal.

By comparing the reflectance map at 750 nm (Figure \ref{fig: orbital-parameters-gusev}a) and the $\omega$ map (Figure \ref{fig: orbital-parameters-gusev}b)
to the dust index map (Figure \ref{fig: orbital-parameters-gusev-composition}),
three photometric regions are distinguishable. The region composed
of ``dark band feature'' (Figure \ref{fig: context-map-gusev}c,
in green color) is correlated to the lowest dust signature (npOx $\simeq$
0.90). This region is correlated to the lowest $\omega$ values ($\omega_{CRISM}\simeq$0.60, $\sigma\leq0.10$). The ``bright-toned'' region in NE of CRISM observation (Figure \ref{fig: context-map-gusev}c,
in red color) shows high dust signature (npOx close to 1) and the highest $\omega$ values ($\omega_{CRISM}\simeq$ 0.80, $\sigma\leq0.10$). The region around
and on the Columbia Hills (``dark linear features'' region) (Figure
\ref{fig: context-map-gusev}c, in yellow color) is correlated to
intermediate dust index values (npOx $\simeq$ 0.95). This likely results from the surface dust,
partly contaminating the substrate, resulting in a mixed signal from dust
and basalt. This area is correlated to intermediate
$\omega$ values ($\omega_{CRISM}\simeq$ 0.65-0.75, $\sigma\leq0.10$).

The single scattering albedo value decreases with the absorption coefficient:
for a given particle size, a high absorption coefficient provides
a low single scattering albedo value. At the studied wavelength (750
nm), similar optical constants ($n$, the refractive index and $k$,
the absorption coefficient) are observed between the typical Martian
dust ($n$=1.50, $k$=0.001, estimated from CRISM observation \citep{wolff2009})
and a typical basalt ($n$=1.52, $k$=0.0011, estimated from laboratory
measurements \citep{Pollack1973}). Consequently, the spatial variation
of the parameter $\omega$, observed in Figure \ref{fig: orbital-parameters-gusev}b,
cannot be explained solely by a composition variation.

The single scattering albedo value increases when the particle size 
decreases. Because dust and basalt have very close optical constants,
the highest $\omega$ values in the ``bright-toned' region can be
explained by the presence of finer particles, compared to the region
of ``dark band feature'', consistent with the in situ observations (Figure \ref{fig: in-situ-observation-gusev}). 
Intermediate $\omega$ values are observed in the region with numerous dust devil
tracks with a width less than 100 meters, which is
lower than the CRISM spatial resolution ($\sim$200m/pixel). As such signal
of some CRISM pixels may be a mixed signal between dust and basaltic sands.
Indeed, along its traverse, the rover crossed albedo boundaries observed
from Pancam images:  low-albedo dust devil tracks (0.20$\pm$0.02)
and high-albedo dustier deposits (0.30$\pm$0.02) \citep{bell2004a}.
From the high resolution MI images
of the soils on the aeolian features, \citet{greeley2004,Greeley2005a} 
provided explanations for these albedo contrasts as follows. Inside the track, 
dust is partly removed and the underlying basalt is observed
as coarse sand grains ($\simeq$ 1 mm in diameter). 
Some dust particles remain resulting in an intimate mixture (Figure
\ref{fig: in-situ-observation-gusev}c, top). Outside the tracks,
a relatively thick dust cover may remain on the top of the basalt
in a stratified mixture (Figure \ref{fig: in-situ-observation-gusev}c,
bottom). In the transition of outside and inside the tracks, a spatial mixture of both 
units may be detected.

The in situ observations from the MI instrument showed that dust particles
are often observed as aggregates of individual unresolved finer particles
\citep{herkenhoff2004a,Sullivan2008,Vaughan2010}, where the subparticles
can act like internal scatterers and increase the single scattering albedo. 

The spatial variation of $\omega$ parameter seems
to be related at first order to the spatial variation of the particle
size. This variation is caused by the dust removal (decrease in the
$\omega$ values) and deposition (increase in the $\omega$ values)
by aeolian processes by strong winds such as storms (such as the ``dark
band feature'' region) or dust devils where silt-sized and sand-sized materials 
can be removed in suspension and by active saltation \citep{greeley2006b}.
The $\omega$ values in the ``dark linear feature'' region result
from a mixture (stratified, intimate, spatial) between basalt and dust
(Figure \ref{fig: in-situ-observation-gusev}c), as is tested by numerical
modeling in subsection \ref{sub:Mixtures}.


\subsubsection*{3.1.3.2. The particle phase function parameters \label{sub:MER Spirit b and c}}

Figures \ref{fig: orbital-parameters-gusev}c and \ref{fig: orbital-parameters-gusev}d
present the mean values of the posterior PDF for asymmetry parameter
($b$) and the backscattering fraction ($c$), respectively. The $c$
parameter values show backscattering behavior ($>0.5$) and the parameter
$b$ values indicate broad scattering lobe ($<0.5$) for all pixels (Figures
\ref{fig: orbital-parameters-gusev}c, \ref{fig: orbital-parameters-gusev}d
and \ref{fig: orbital-parameters-gusev}f). The local differences
in the map are not significant in comparison to the error ($\sigma<0.2$).

By comparing to the in situ photometric results, the CRISM results
are consistent with $b$ and $c$ values, estimated for Soil unit at the Landing site and the Bonneville Rim sites 
\citep{Johnson2006a} (Figure \ref{fig: in-situ-parameters-gusev}c). 

To provide a physical meaning of the phase function parameters, experimental
studies were conducted on well-characterized natural and artificial
materials \citep{mcGuire1995,shepard2007,souchon2011,Johnson2013}.
\citet{mcGuire1995} studied the scattering properties of different
isolated artificial particles which had different structure types.
Their study showed that smooth clear particles exhibit greater forward
scattering (low values of $c$) and narrower scattering lobes (high
values of $b$), whereas particles characterized by their roughness
or internal scatterers exhibit greater backward scattering (high values
of $c$) and broader scattering lobe (low values of $b$) (Figure
\ref{fig: orbital-parameters-gusev}f). By comparing to the \citet{mcGuire1995}'s
study, the scattering properties acquired from the CRISM observation are
closer to the scattering properties of artificial materials composed
of particles with moderate density of internal scatterers (Figure
\ref{fig: orbital-parameters-gusev}f). Similarly, \citet{souchon2011}
measured, for a comprehensive set of geometries, the reflectance factor
of natural granular surfaces composed of volcanic materials differing
by their grain size, shapes, surface aspect, and mineralogy (including
glass and minerals). By comparing to the \citet{souchon2011}'s work
on natural volcanic samples, the scattering properties of ``dark
band'' region where basaltic sands are exposed (Figure
\ref{fig: context-map-gusev}c, in green color), are closer to the
scattering behavior of pyroclastics from Towada T6 (425-1000 $\mu m$), 
characterized by rounded grains with rough, hollowed and opaque surfaces, 
with facets of phenocrysts and some isolated crystals. The results
show that the grains at the Gusev plain are consistent with rounded grains
which are transported in long distance by wind. Second, the results show
that the grains are composed of a high density of internal scatterers, 
suggesting the presence of either high density of crystals and/or the 
presence of impurities (e.g., bubbles) and/or fractures.

The MI images focused on a typical basaltic soil of coarse grains
and cleaned of dust (Figure \ref{fig: in-situ-observation-gusev}c,
top) showed rounded and relatively spherical sands and granules, as
shown by \citet{greeley2006b}. Moreover, the basaltic materials showed a complex and 
heterogeneous structure created by different crystals (e.g., olivine, pyroxene, plagioclase) 
that may cause the observed backward scattering behavior. Those results are consistent with the $b$
and $c$ values estimated from the CRISM observation. 

\subsubsection*{3.1.3.3. The surface macroscopic roughness parameter\label{sub:MER Spirit theta}}

Figure \ref{fig: orbital-parameters-gusev}e represents the map of
the macroscopic roughness parameter ($\bar{\theta}$) values. High
$\bar{\theta}$ values ($\bar{\theta}$ $\sim$15-20$^\circ$,
$\sigma\leq$5$^\circ$) are observed.

To explain the high $\bar{\theta}$ values, the mean slope was calculated
from the HiRISE DTM at scale of 1 meter per pixel (Figure \ref{fig: context-map-gusev}d).
In the cratered plains, we can identify 0.1-1 km diameter-size craters
characterized by the highest mean slope (mean slope: 5 - 10$^\circ$).
The intercrater plain is characterized by the lowest mean slope (mean
slope: $<$ 5$^\circ$) (Figure \ref{fig: context-map-gusev}d).
By comparing the $\bar{\theta}$ parameter map to the mean slope,
we note that the high $\bar{\theta}$ values are not correlated with
a high mean slope. This suggests that the macroscopic roughness parameter
is not representative of the local topography at the meter- scale.
The $\bar{\theta}$ parameter is more sensitive to the microscale (from the particle size to a few millimeters) 
and thus to the grain organization, as suggested by \citet{Shepard1998,Helfenstein1999,cord2003,shkuratov2005,Shaw2013}.

Average surface macroscopic roughness values, obtained from CRISM data ($\sim$13$^\circ$ $\pm$ 4$^\circ$ 
Figure \ref{fig: orbital-parameters-gusev}e) are close to those estimated from in situ photometric observations 
($\sim$14$^\circ$ $\pm$ 2$^\circ$, Figure \ref{fig: in-situ-parameters-gusev}b) from 
Pancam \citep{Johnson2006a}. This suggests that the macroscopic roughness, estimated 
from CRISM data, is more representative of the microscale of the soil unit 
in agreement with earlier findings based on HRSC data \citep{Jehl2008}. 

\begin{figure}[H]
\begin{centering}
\includegraphics[scale=0.50]{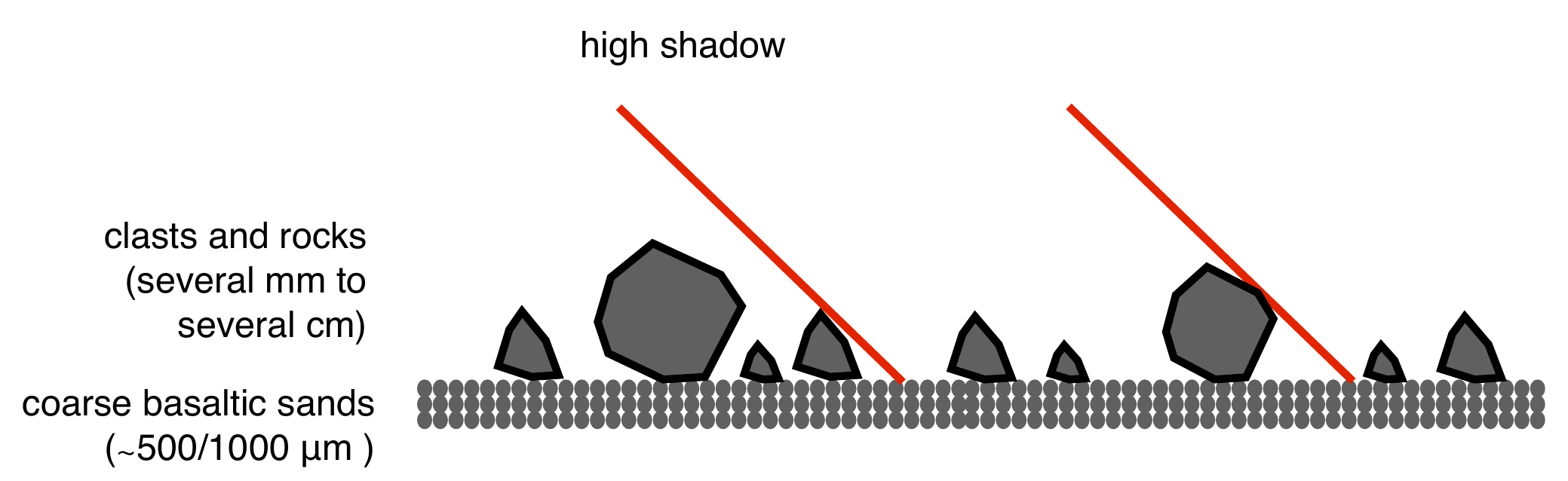}
\par\end{centering}

\caption{Schematic representation of the meaning of the macroscopic roughness values and the relation to the 
surface roughness in the Gusev plain. ($\sim500$ $\mu m$ for basaltic sands and several 
millimeters to several hundreds centimeters for the clastes and rocks). \label{fig: roughness - gusev}}

\end{figure}

The HiRISE image may also provide information about the surface morphology and 
the surface roughness. The HiRISE image (Figure \ref{fig: context-map-gusev}b) shows numerous
craters which may be accompanied by angular mm- to cm-sized ejected clasts and rocks. The high population
of such ejecta materials having an angular shape may create a high shadow hiding, leading
to a high macroscopic roughness values (Figure \ref{fig: roughness - gusev}). 


\subsection{MER-Opportunity landing site at Meridiani Planum \label{sub:MER Opportunity}}

\subsubsection{Selection of CRISM observations \label{sub:MER Opportunity selection }}

Up to seven FRT observations are available at MER-Opportunity landing
site from the beginning of the CRISM operations to september 2010.
Among all, the observations FRT\#B6B5 and FRT\#193AB show the best 
statistic results, relative to the MARS-ReCO procedure, 
and the Bayesian inversion (satisfactory values of the non-uniform
($k$), the standard deviation ($\sigma$) criteria and the bimodality
of the single scattering albedo PDF criterion). However, the statistic results for the parameter
$c$ of the observation FRT\#B6B5 show a number of pixels with
$\sigma_{c}\leq0.20$ equal to 8\%. To improve the geometric diversity
of the FRT\#B6B5, we combined the observations FRT\#B6B5 and FRT\#334D 
because they are superimposed, have complementary geometric acquisitions,
and no obvious surface changes were observed. After the combination, the statistics
of Bayesian inversion are improved, especially for the parameter $c$, 
showing a number of pixels with $\sigma_{c}\leq0.20$ equal to 25\%.
The solutions, estimated from the combination of FRT\#B6B5 and FRT\#334D
and from FRT\#193AB, are presented in this work (Table \ref{tab:Selected CRISM FRT}).

\subsubsection{Geological context and study areas}

The MER-Opportunity landed in the Plains, Hematite-bearing
(Ph) unit composed of patches light-toned sedimentary materials (etched
terrain, ET2), overlain by a unconsolidated veneer of crystalline
hematite bearing (plains mantle, Pm), detected by the TES instrument \citep{Christensen2000,Christensen2001}.

\begin{figure}[H]
\begin{centering}
\includegraphics[scale=0.55]{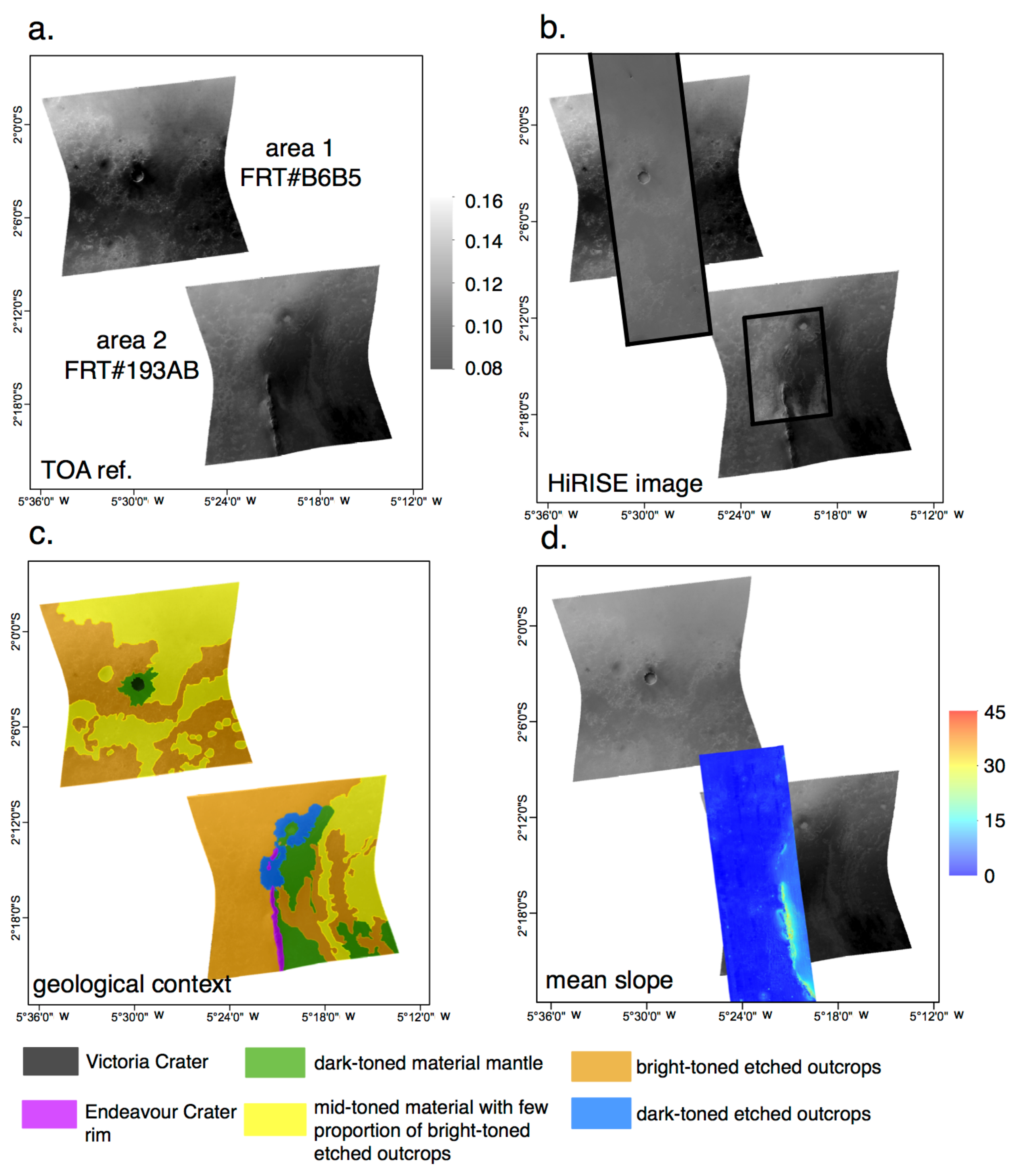}
\par\end{centering}

\caption{Geological context and studied area: a. CRISM TOA central images (FRT\#B6B5
and FRT\#193AB) at 20m/pxl at 750 nm, b. CRISM TOA central images
overlapped by the associated HiRISE image (PSP\_009141\_1780\_RED
and ESP\_032573\_1775\_RED), c. CRISM TOA overlapped by the geological
context map showing the different units and structures, d. CRISM TOA
overlapped by the mean slope map at 1m derived from HiRISE Digital
Terrain Model (DTM) (DTEEC\_018701\_1775\_018846\_1775\_U01).\label{fig: context-map-meridiani}}
\end{figure}

Two sites are presented in this work: the surrounding of the Victoria
crater (results from the combination of FRT\#B6B5 and FRT\#334D observations,
called area 1), and the West side of Endeavour crater rim (results
from the FRT\#193AB observation, called area 2) (Figure \ref{fig: context-map-meridiani}a).


\paragraph*{Victoria crater (area 1)}

The HiRISE image, taken at the same local time as the FRT\#B6B5, was 
used to make the geological context map. Four geological units are 
discernible in the plains: 

(1) Victoria crater (Figure \ref{fig: context-map-meridiani}c,
black color unit),

(2) dark-toned materials mantle, around the Victoria crater (Figure
\ref{fig: context-map-meridiani}c, green color unit), 

(3) no or low extended areas of bright-toned materials in the aeolian
ripple troughs, corresponding to the bedrock (etched terrain) associated
with mid-toned materials in the ripple crests (Figure
\ref{fig: context-map-meridiani}c, yellow color unit),

(4) high extended areas of bright-toned materials in the aeolian ripple
troughs, corresponding to the bedrock (etched terrain), associated with
mid-toned materials in the ripple crests (Figure
\ref{fig: context-map-meridiani}c, orange color unit). 

Orbital reflectance spectra acquired by OMEGA on-board MEx, over the
MER-Opportunity landing site, are dominated by a basaltic sand cover
mixed with crystalline hematite and small amount of dust \citep{arvidson2006b}. 
The abundant hydroxylated and hydrated sulfate minerals, observed by
the rover \citep{Arvidson2011,Squyres2006}, corresponding to the brighter
terrains (Figure \ref{fig: context-map-meridiani}c,
red and orange color units) are identified in high resolution images
(Figure \ref{fig: context-map-meridiani}b) but have not been detected
over the landing site at the OMEGA and CRISM spatial resolution scales.
The region dominated by ripples plains is characterized by low THEMIS TI values ($\sim$140 -
145 $J.m^{-2}.K^{-1}.s^{-1/2}$ \citep{Arvidson2011}), suggestive
of the presence of fine-grained aeolian materials. The region where bedrock is visible, the THEMIS TI values are slightly
higher than ripples, suggestive of the presence of indurated materials
\citep{Arvidson2011}. 

\begin{figure}[H]
\begin{centering}
\includegraphics[scale=0.40]{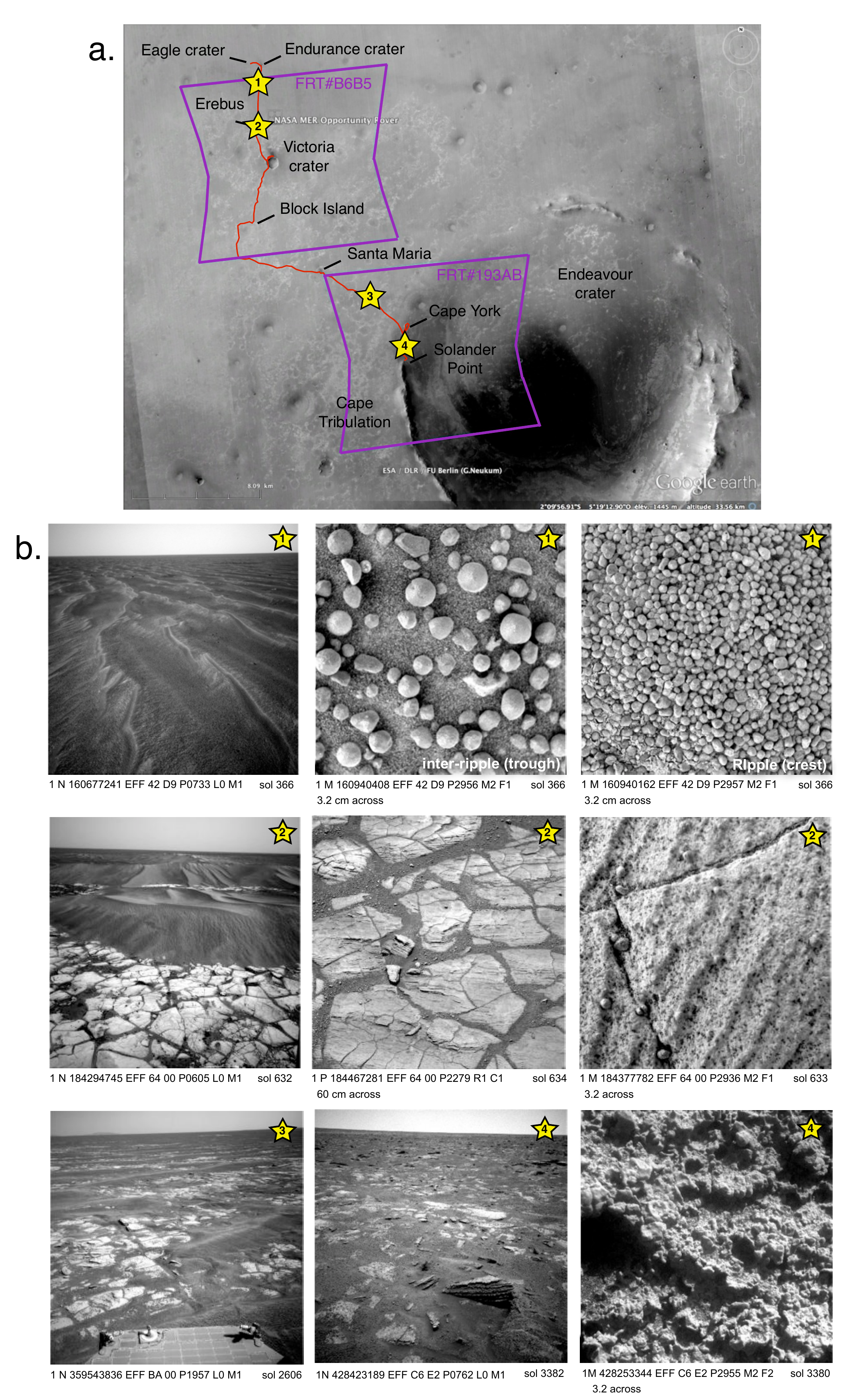}
\par\end{centering}

\caption{In situ observations from Navcam/MER (N), Pancam/MER (P) and MI/MER
(M). (a) location of FRT\#B6B5 (FRT\#334D has the same footprint as
FRT\#B6B5) and FRT\#193AB CRISM footprints (in purple color), the
rover traverse (in red color) and the location of the selected in
situ observations (star), (b) Navcam, Pancam and MI observations of
the aeolian ripple region composed of hematitic concretions, basaltic
sands and a small amount of dust (top row), Navcam, Pancam and MI
observations of the aeolian ripple region with sulfate-rich outcrops
at Erebus area (middle row) and Navcam and MI observations dark-toned sulfate-rich outcrops 
at Botany Bay at the west part of the Endeavour crater (bottom row). \label{fig: in-situ-observation-meridiani}}
\end{figure}

\citet{Squyres2006} and \citet{Arvidson2011} provided an overview
of key observations of soils and rocks observed by MER-Opportunity. 
Along the rover traverse (Figure \ref{fig: in-situ-observation-meridiani}a), the
in situ observations showed that the Meridiani plain surfaces are
covered by aeolian ripples (Figure \ref{fig: in-situ-observation-meridiani}b,
top). The soil in the ripple troughs is dominated by rounded hematitic
concretions, and fragments and their size is larger than those found
in ripple crest soils (Figure \ref{fig: in-situ-observation-meridiani}), 
whereas the soil in the ripple crest is dominated by well-sorted hematitic
concretions of a relatively uniform size distribution (Figure \ref{fig: in-situ-observation-meridiani})
(granule : 1-2 mm in diameter). A few millimeters beneath the lag of
deposit, the interior is dominated by a mixture of basaltic sands,
fragments of hematitic concretions and dust \citep{arvidson2006a,herkenhoff2004b,Herkenhoff2006,Jerolmack2006,Soderblom2004b,weitz2006,Squyres2004,Squyres2006,Sullivan2005}
(Figure \ref{fig: in-situ-observation-meridiani}b, top row). From
the Erebus Crater to the Endeavour crater (Figure \ref{fig: in-situ-observation-meridiani}a),
bright-toned flat-lying bedrock outcrops (Figure \ref{fig: context-map-meridiani}c,
in orange color units) are visible and are exposed in ripple troughs, 
underlying the thin plain soil. The bedrocks are characterized 
by fine laminations, have a high concentration
of sulfur and contain abundant sulfate salts (Figure \ref{fig: in-situ-observation-meridiani}b,
middle line) \citep{Squyres2004}. MI images showed that the outcrops
bedrocks are composed of: (i) moderate rounded well-sorted sand grains
(from 0.2 to 1 mm) forming mm-scale laminations, (ii) fine-grained
and coarser precipitated cement crystals, (iii) cm-sized vugs that
record the early diagenetic growth and subsequent dissolution of crystals,
and (iv) 3- to 5-mm sized hematitic spherules embedded within the
outcrops \citep{Herkenhoff2006,Squyres2004c}.

\paragraph*{West side of Endeavour crater rim (area 2)}

The HiRISE image, taken at the same local time as the FRT\#193AB, was
used to make the geological context map.

At the west part of the rim in the Meridiani plain, the region is dominated by
patches of bright-toned materials in the aeolian ripple troughs, corresponding
to the bedrock (etched terrain), associated with mid-toned materials
in the ripple crests (Figure \ref{fig: context-map-meridiani}c, orange
color unit). This unit is the same as the etched terrains observed in the plains
around the Victoria Crater. However, the unit is characterized by higher THEMIS TI values 
($\sim$155-180 $J.m^{-2}.K^{-1}.s^{-1/2}$ \citep{Arvidson2011}) in this area. The mean slope values
derived from HiRISE DTM show low slopes, less than 5$^\circ$ (Figure
\ref{fig: context-map-meridiani}d).

The region around the Endeavour Crater rim, including the Botany
Bay area around the Cape York region (Figure \ref{fig: context-map-meridiani}c, blue color unit),
is composed of dark-toned outcrops (etched terrain),
visible in the HiRISE image (Figure \ref{fig: context-map-meridiani}b), 
detected to be hydrated sulfate-rich bedrocks from
CRISM spectra \citep{Wray2009}, and corresponding to the Burns formation sulfate-rich sandstones \citep{Arvidson2014}.

The discontinuous Endeavour rim (Cape York in the north and Cape
Tribulation in the south) (Figure \ref{fig: context-map-meridiani}c,
purple color unit) shows the highest THEMIS TI values
($>$340 $J.m^{-2}.K^{-1}.s^{-1/2}$ \citep{Chojnacki2010}),
suggestive of the presence of indurated materials \citep{Arvidson2011,Chojnacki2010}.
The rim exposes basalt and iron and magnesium-rich smectite clay minerals
\citep{Wray2009}. The mean slope values, derived from HiRISE DTM, show
high slopes, greater than 15$^\circ$ (Figure \ref{fig: context-map-meridiani}d).

The Endeavour crater floor is composed of several units:

(1) a mantle of dark-toned materials (Figure \ref{fig: context-map-meridiani}c, 
green color unit), similar to the Ph unit \citep{Chojnacki2010}.

(2) low extended areas of bright-toned materials, corresponding to the bedrock, similar to the etched unit,
associated with mid-toned materials in the ripple crests (Figure
\ref{fig: context-map-meridiani}c, yellow color unit).

(3) high extended areas of bright-toned materials, corresponding to the
bedrock, similar to the etched unit associated with mid-toned materials
in the ripple crests (Figure \ref{fig: context-map-meridiani}c,
orange color unit).

The in situ observations showed bright-toned outcrops
from the Erebus Crater to the Endeavour crater (Figure \ref{fig: in-situ-observation-meridiani}a)
which are associated with sulfate-rich bedrocks exposed in ripple troughs (Figure \ref{fig: in-situ-observation-meridiani}b, star 3).
In the Botany Bay area, in situ observations showed flat polygonal
fractured outcrops. The rocks are composed of layered features, with
rounded coarse-grained materials (sand-sized particles), embedded in
a matrix and cement, with a thin covering of soil \citep{Arvidson2014} (Figure \ref{fig: in-situ-observation-meridiani}b, star 4).


\subsubsection{Surface material photometric parameters \label{sub:MER Opportunity parameters}}

\subsubsection*{3.2.3.1. The single scattering albedo parameter \label{sub:MER Opportunity w}}

\begin{figure}[H]
\begin{centering}
\includegraphics[scale=0.55]{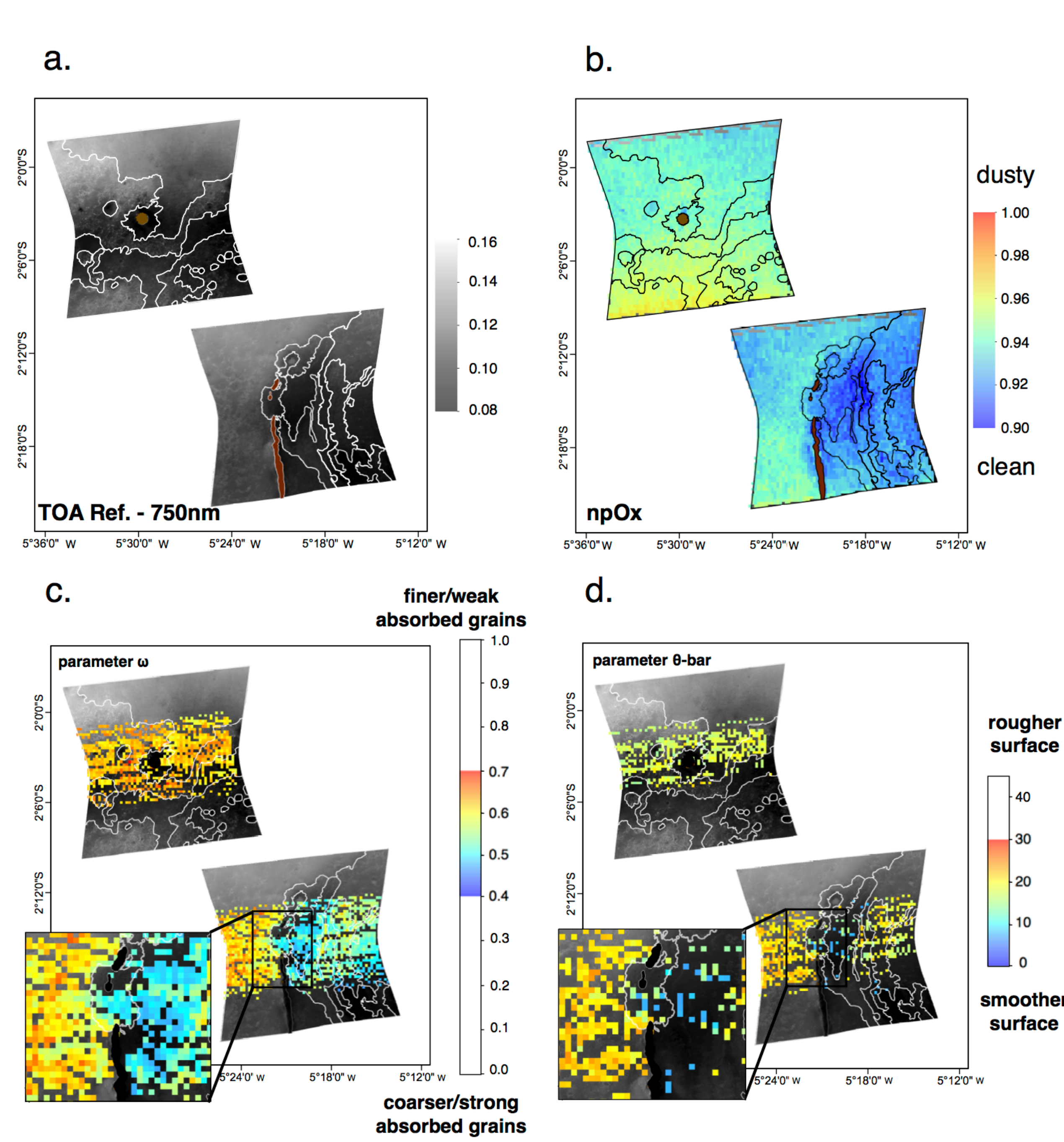}
\par\end{centering}

\caption{CRISM mapping of the combination of FRT\#B6B5 - FRT\#334D and FRT\#193AB.
a. TOA reflectance map in I/F unit at 20m/pixel at 750 nm. b. map
of npOx spectral index (dust index) from FRT\#B6B5 and from FRT\#193AB
at 200m/pixel. c. parameter $\omega$ map at 200m/pixel ($\sigma_{\omega}\leq0.10$).
d. parameter $\bar{\theta}$ map at 200/pixel ($\sigma_{\bar{\theta}}\leq$5$^\circ$).
The colored pixels correspond to the value of the mean PDF at about
200m/pixel. Only the middle part of the central image is covered with
all additional geometric images (up to 22 images for area 1 and up
to 11 images for area 2) that it is why the photometric results are
obtained in this area. Regions dominated by abrupt and high local topography 
(Victoria crater and Endeavour crater rim) are masked off in the photometric maps that would otherwise make  
the atmospheric correction and the photometric study more 
challenging (section 2.3). \label{fig: orbital-parameters-meridiani-1}}
\end{figure}

Figure \ref{fig: orbital-parameters-meridiani-1}c 
represents the map of the single scattering albedo values ($\omega$).
Spatial variations of the $\omega$ parameter are observed at 200m/pixel
scale. In area 1, the regions mainly composed of
dark-toned (Figure \ref{fig: context-map-meridiani}c, green color
unit) and of mid-toned materials (Figure \ref{fig: context-map-meridiani}c,
yellow color unit) are associated with the lowest $\omega$ values
($\omega_{CRISM}$ $\leq0.60$, $\sigma\leq0.10$). The regions with
a high proportion of bright-toned outcrops and accompanied with mid-toned
materials (Figure \ref{fig: context-map-meridiani}c, orange color
unit) are associated with the highest $\omega$ values ($\omega_{CRISM}$
$\simeq0.60-0.65$, $\sigma\leq0.10$). In area 2, the crater floor
composed of dark-toned material mantle (Figure \ref{fig: context-map-meridiani}c,
green color unit) and the Botany Bay region (Figure \ref{fig: context-map-meridiani}c,
blue color unit) are associated with the lowest $\omega$ values ($\omega_{CRISM}$
$\simeq0.40-0.50$, $\sigma\leq0.10$). The bright-toned materials
observed in the crater floor (Figure \ref{fig: context-map-meridiani}c,
orange color unit) are associated with high $\omega$, but lower than 
the bright-toned outcrops, observed in area 1 and in
area 2 (Figure \ref{fig: context-map-meridiani}c, orange color unit)
($\omega_{CRISM}$ $\simeq0.60$, $\sigma\leq0.10$).


\begin{figure}[H]
\begin{centering}
\includegraphics[scale=0.42]{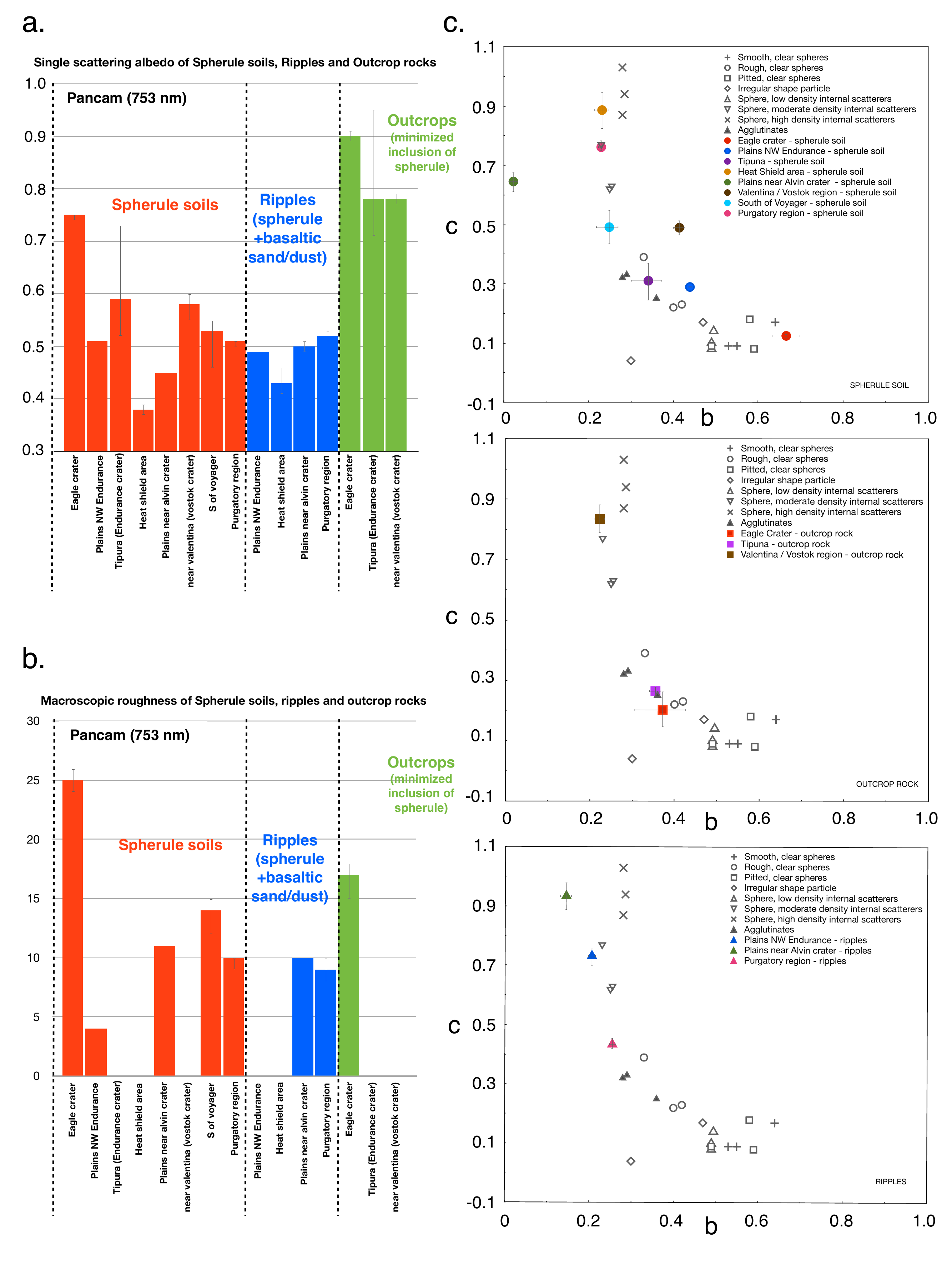}
\par\end{centering}

\caption{In situ photometric results from Pancam/MER: Mean and uncertainties
of a. the single scattering albedo ($\omega$), b. the macroscopic
roughness ($\bar{\theta}$), c. the particle phase function parameters
($b$ and $c$) overplot to the experimental $b$ and $c$ values
pertaining to artificial particles measured by \citet{mcGuire1995}.
All photometric parameters are estimated at 753 nm for different
geological units at different sites along the traverse (from Sol 11
to Sol 473) \citep{Johnson2006b}.  \label{fig: in-situ-parameters-meridiani}}
\end{figure}


From orbital observations, four surface components are observed: (i) hematitic
concretions, (ii) basaltic sands, (iii) dust and (iv) sulfate, consistent
with the in situ observations. The optical constants ($n$, and $k$), 
at the studied wavelength (750 nm) shown in Table 3, are 
$n$=1.52, $k$=0.0011 for basalt \citep{Pollack1973},
$n$=1.50, $k$=0.001 for dust \citep{wolff2009}, $n$=2.805,
$k$=0.03478 \citep{Sokolik1999} for hematite and $n$=1.5, $k$=0.00001 for sulfate
\citep{Roush2007}. Hematite has the highest $k$ coefficient (leading to lowest $\omega$) and
the sulfate mineral the lowest (leading to highest
$\omega$). These are consistent with the CRISM $\omega$ results, 
where the lowest $\omega$ values are observed in aeolian ripples, mainly composed of hematitic concretions
(Figure \ref{fig: context-map-meridiani}c, yellow
and green color unit). The highest $\omega$ values are observed
in regions associated with sulfate-rich bright-toned outcrops 
(Figure \ref{fig: context-map-meridiani}c, red and orange color units).

In situ images from Pancam and MI provide constraints on the particle
size of each component, helpful for the $\omega$ values interpretation:
(i) the basaltic granular materials are characterized by fine-grained sands with a mean size around
50 to 150 $\mu m$ \citep{Herkenhoff2006,herkenhoff2004b,weitz2006}
(Figure \ref{fig: in-situ-observation-meridiani}b, top row), (ii)
the hematitic concretions are characterized with a particle size around 1-2 mm \citep{Herkenhoff2006,weitz2006,herkenhoff2004b}
(Figure \ref{fig: in-situ-observation-meridiani}b, top row), (iii) the sulfate-rich bright-toned rocks are composed of moderate rounded well-sorted sand grains (from 0.2 to 1 mm) forming mm-scale laminations, fine-grained and coarser cement crystals, cm-sized vugs, and 3- to 5-mm sized hematitic 
spherules embedded within the outcrops \citep{Herkenhoff2006,Squyres2004c}
(Figure \ref{fig: in-situ-observation-meridiani}b, middle row),
(iv) dust deposit of silt size ($<$4 $\mu m$ in diameter) 
\citep{lemmon2004} and (v) sulfate-rich dark-toned rocks 
composed of coarser grains (sand-sized) \citep{Arvidson2014}
than the sulfate-rich bright-toned rocks (fine-sized).
The coarse hematitic concretions (mm-sized) and the coarse grains
in the sulfate-rich dark-toned rocks accentuates the decrease of $\omega$
values. The fine particles that compose the bright-toned outcrop and the dust cover in 
the plains accentuates the increase of $\omega$ values, consistent with the CRISM $\omega$
estimates.

For the hematite-rich concretions, the MI images did not show macroscopic
evidence of internal structures within the $\sim$100 $\mu m$ resolution
of the instrument \citep{herkenhoff2004b,McLennan2005,Herkenhoff2008}.
However, the emissivity measurements of the hematitic concretions
from TES on-board MGS and Mini-TES on-board MER-Opportunity suggested
that the emissivity is dominated by emission along the crystallographic
$c$ axis, explained by the lack of a 390 $cm^{-1}$ feature in the
hematite-rich spherule spectra \citep{Lane2002,Glotch2004}. To explain
this observation, \citet{Glotch2006} from modeling and \citet{Golden2008}
from experimental study suggested the presence of a high density of
internal structure for the hematitic spherules (randomly oriented
platy hematite crystal, concentric growth, or fibrous growth along
the radial direction). For the sulfate-rich rocks, the MI images (Figure
\ref{fig: in-situ-observation-meridiani}b) showed macroscopic evidence
of heterogeneity in the sedimentary rocks with the presence of fine-grained
and coarser cement crystals. For the basaltic granular medium, the MI resolution did not
show evidence of an internal structure within the $\sim$100 $\mu m$
resolution of the instrument. However, the basaltic grains are generally composed
of particles with high internal structures, such as minerals or bubbles,
similar to the Gusev basaltic sands. For the three components (hematitic
concretions, sulfate-rich rocks and basalt), the $\omega$ values
may be higher than their counterpart (those with no internal structures), 
consistent with the CRISM $\omega$ estimates.

The single scattering albedo values, estimated by
CRISM, are compared to those estimated from Pancam measurements \citep{Johnson2006b}
(Figure \ref{fig: in-situ-parameters-meridiani}a). Indeed,
the in situ instruments can distinguish rocks and soils whereas CRISM
observes an extended area including rocks and soils (unit mixtures).
Pancam measurements were taken at 753 nm for different geological
units (rocks and soils) \citep{Johnson2006b}: (i) the Outcrop unit refers to as the sulfate-rich
bedrock, (ii) the Spherule soil unit refers to as unconsolidated materials
of basaltic sands and of hematitic concretions, and (iii) the Ripple
soil unit refers to as the ripple crest materials composed of basaltic
sands, hematitic concretions and a small amount of dust (the population
of hematitic concretions is greater than the Spherule soil unit).
The Outcrop unit showed the highest $\omega$ values ($\omega_{Pancam}<0.90$).
For the Ripple soil and Spherule soil units, the lowest $\omega$
values were modeled ($\omega_{Pancam}<0.60$) (Figure \ref{fig: in-situ-parameters-meridiani}a). However, the Ripple
soils unit showed lower $\omega$ values ($\omega_{Pancam}<0.50$)
than the Spherule soils ($\omega_{Pancam}<0.60$)( Figure \ref{fig: in-situ-parameters-meridiani}a). This discrepancy
can be explained by the higher abundance of hematitic concretions
in the ripple crests (accumulation during the wind transport) compared
the ripple troughs and the soils \citep{herkenhoff2004b,Herkenhoff2006,weitz2006}
(Figure \ref{fig: in-situ-observation-meridiani}b, top row). The direct 
comparison of the CRISM results to those estimated from in situ Pancam
measurements is presented below.

(1) The regions mainly composed of mid-toned materials
in areas 1 and 2 (Figure \ref{fig: context-map-meridiani}c, yellow
color unit) show consistent values ($\omega_{CRISM}$ $\leq0.60$,
$\sigma\leq0.10$) with the Pancam estimates, obtained for the Spherule
soil and Ripple soil units ($\omega_{Pancam}$ $<0.60$).

(2) The regions with a high proportion of bright-toned
outcrops and accompanied with mid-toned materials in areas 1 and 2
(Figure \ref{fig: context-map-meridiani}c, orange color unit) show
lower values ($\omega_{CRISM}$ $\simeq0.60-0.70$, $\sigma\leq0.10$)
than the Pancam estimates ($\omega_{Pancam}$ $\simeq0.9$), which
can be explained by the spatial resolution difference between Pancam
and CRISM. Indeed, Pancam distinguishes individual units of rock (e.g., Outcrop unit) and soil
(e.g., Spherule soil and Ripple soil) whereas CRISM measures areas
integrating different geological units. A mixture of Spherule soils
or Ripple soil units, characterized by lower $\omega$ values ($\omega_{Pancam}<0.6$),
with the Outcrop unit, characterized by higher $\omega$ values 
($0.77<$ $\omega_{Pancam}$ $<0.90$), seems to slightly decrease 
the $\omega$ values, consistent with the lower $\omega$ values obtained from CRISM observation. In their experimental
studies, \citet{Johnson2013} showed that the addition of 2-3 mm-
sized spherules with thin hematite rims, on an analog of the sulfate-rich
sedimentary rocks, composed of finer particles, decrease the $\omega$
values. The influence of mixtures in the scattering behavior between
hematitic concretions and basalt and between hematitic concretions
and sulfate are tested by a numerical modeling in Subsection \ref{sub:Mixtures}.

(3) The regions including the Botany Bay area, composed of dark-toned
sulfate-rich outcrops (Figure \ref{fig: context-map-meridiani}c,
blue color unit), show lower values ($\omega_{CRISM}$ $\simeq0.45-0.50$,
$\sigma\leq0.10$) than those estimated for the Outcrops unit by Pancam
($0.77<$ $\omega_{Pancam}$ $<0.9$). This suggests that there is a difference
in the physical properties. Coarser grains can explain the low $\omega$ values in the dark-toned outcrops
regions, confirmed by the in situ observations \citep{Arvidson2014} as could the 
prevalence of dark-toned basaltic grains and coatings interspersed with the outcrop 
materials within the Botany Bay area.

(4) The regions composed of a dark-toned mantle
in the Endeavour crater floor (Figure \ref{fig: context-map-meridiani}c,
green color unit in area 2) show lower values ($\omega_{CRISM}$ $\simeq0.40-0.50$,
$\sigma\leq0.10$) than those estimated from the Spherule soil and
Ripple soil units ($\omega_{Pancam}$ $<0.60$) and than those estimated
at the region around Victoria Crater (Figure \ref{fig: context-map-meridiani}c,
green color unit in area 1). Lower values are also observed compared
to the region of aeolian ripples without sulfate-rich outcrops (Figure
\ref{fig: context-map-meridiani}c, in yellow color unit). The npOx
index value (Figure \ref{fig: orbital-parameters-meridiani-1}b) indicates
that the surface is less contaminated by dust, compared
to all units in area 1. As observed at MER-Spirit landing site (Subsection
\ref{sub:MER Spirit w}), dust is characterized by high $\omega$
values ($\omega_{CRISM}\sim0.80$). Consequently, its relative paucity in
the crater floor is consistent with the lower mean $\omega$ values.

\subsubsection*{3.2.3.2. The particle phase function parameters \label{sub:MER Opportunity b and c}}

\begin{figure}[H]
\begin{centering}
\includegraphics[scale=0.50]{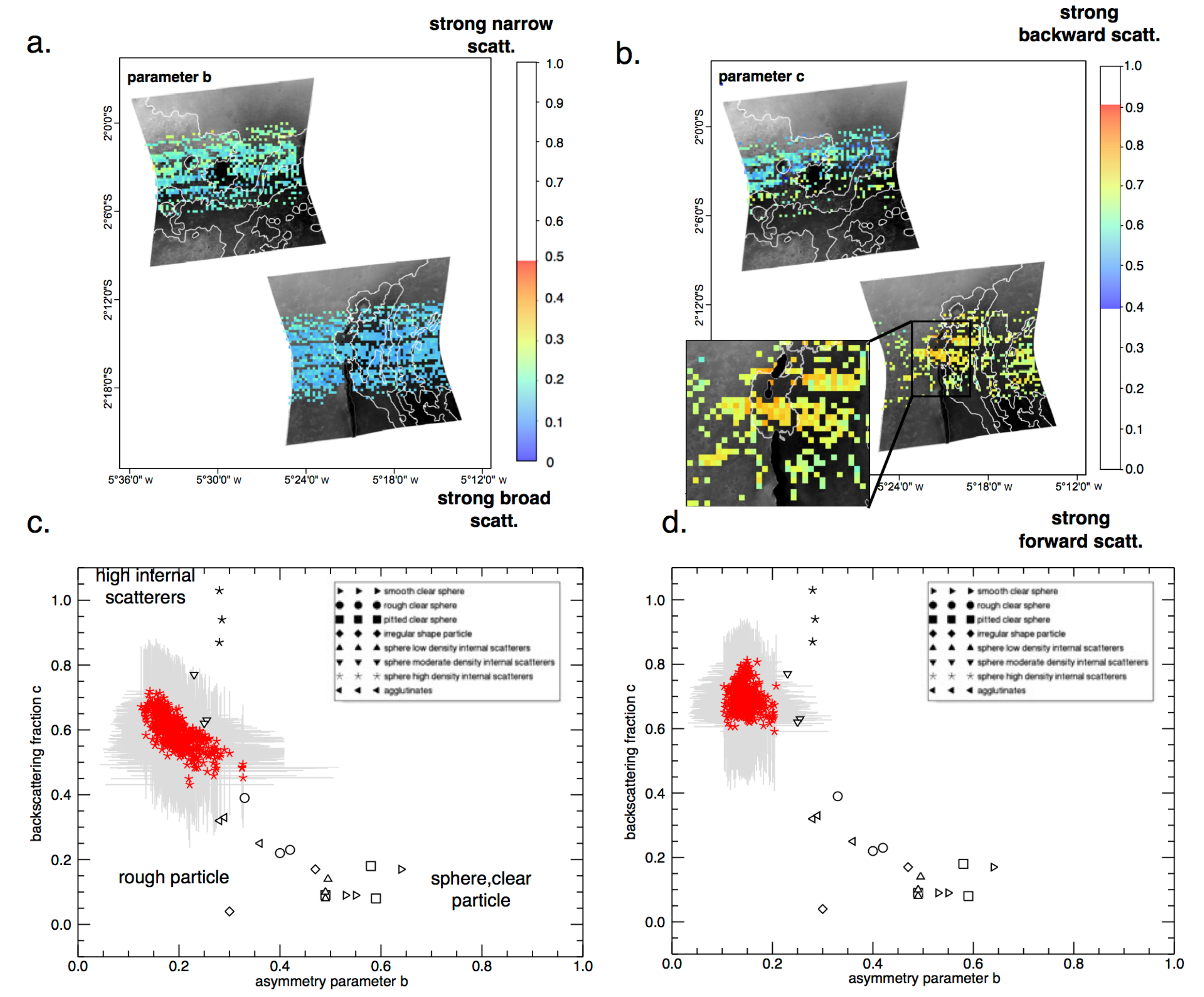}
\par\end{centering}

\caption{CRISM mapping of the combination of FRT\#B6B5 - FRT\#334D and FRT\#193AB.
a. parameter $b$ map at 200m/pixel ($\sigma_{b}\leq0.20$). b. parameter
$c$ map at 200m/pixel ($\sigma_{c}\leq0.20$). The colored pixels
correspond to the value of the mean PDF at about 200m/pixel. Only
the middle part of the central image is covered with all additional
geometric images (up to 22 images for area 1 and up to 11 images for
area 2) that it is why the photometric results are obtained in this
area. c. and d. graph of the asymmetry parameter ($b$) versus backscattering
fraction ($c$) estimated from FRT\#B6B5 - FRT\#334D (Figure 11c) and from FRT\#193AB (Figure 11d) 
plotted with experimental values on artificial particles 
from \citet{mcGuire1995}. \label{fig: orbital-parameters-meridiani-2}}.
\end{figure}

Figures \ref{fig: orbital-parameters-meridiani-2}a and \ref{fig: orbital-parameters-meridiani-2}b
present the asymmetry parameter ($b$) and the backscattering fraction
($c$) respectively. At first order, the surface material is characterized
by a broad backscattering behavior ($c>0.5$ and $b<0.5$ similar to particles with moderate density of internal
scatterers (Figure \ref{fig: orbital-parameters-meridiani-2}c)). Moreover, we note that the $c$ values are slightly higher in
area 2 than in the area 1 (Figure \ref{fig: orbital-parameters-meridiani-2}b).


The particle phase function parameters estimated by CRISM are
compared to those estimated from Pancam measurements at 753 nm \citep{Johnson2006b}
(Figure \ref{fig: in-situ-parameters-meridiani}c) to understand the
mean surface scattering behavior.

(1) The crater floor composed of dark-toned material
mantle (Figure \ref{fig: context-map-meridiani}c, green color unit
in area 2) shows the highest $c$ values ($c_{CRISM}\sim0.75-0.80$,
$\sigma\leq0.20$). Values are consistent with those estimated
for Spherule soil and Ripple soil units (Figure \ref{fig: in-situ-parameters-meridiani}c). As discussed previously, 
the $b$ and $c$ values are consistent with the presence of an internal structure in the hematitic concretions, determined 
from Mini-TES spectra. Moreover, the $b$ and $c$ values are consistent with the physical properties
of typical basaltic materials, composed of high internal structure such as the basaltic sands
at MER-Spirit area (Subsection \ref{sub:MER Spirit b and c}).

(2) The dark-toned sulfate-rich bedrocks (e.g., Botany Bay) (Figure \ref{fig: context-map-meridiani}c, blue color
unit in area 2) show the highest $c$ values ($c_{CRISM}\sim0.75-0.80$,
$\sigma\leq0.20$). We note a difference in the $c$ values compared
to aeolian ripples with sulfate-rich outcrops (Figure \ref{fig: context-map-meridiani}c
orange and red color units), that suggests a difference in physical
properties. The high $c$ values can be explained by the presence
of greater heterogeneity inside the outcrops, compared to the bright-toned outcrops.
This is consistent with the in situ observations which showed that
the dark-toned sulfate rocks are composed of coarser grains (sand-sized
grains), embedded in a matrix and cement \citep{Arvidson2014}).

(3) The regions mainly composed of mid-toned materials
(Figure \ref{fig: context-map-meridiani}c, yellow color unit in areas
1 and 2), show  intermediate $c$ values ($c_{CRISM}\sim0.60-0.75$,
$\sigma\leq0.20$).  Values are consistent with those estimated for the Spherule soil and Ripple soil units from
Pancam,  (Figure \ref{fig: in-situ-parameters-meridiani}c). Lower $c$ values are
obtained, compared to the dark-toned mantle at Endeavour crater floor
(Figure \ref{fig: context-map-meridiani}c, green color unit). Again, this
difference can be explained by the presence of small amount of dust
in the aeolian ripples of dark-toned materials (the npOx index is
close to 0.94, Figure \ref{fig: orbital-parameters-meridiani-1}b).
\citet{johnson2006c} showed a more forward scattering behavior for
isolated dust particles, consistent with the scattering behavior of
typical atmospheric dust particles, as modeled by \citet{Tomasko1999}
and \citet{lemmon2004}, over the Mars Pathfinder and MER landing sites.
The mixture of dust with the unconsolidated materials of hematitic
concretions and basaltic sands can explain the decrease of $c$ values.

(4) The regions with a high proportion of bright-toned
outcrops and accompanied with mid-toned materials in area 1 (Figure
\ref{fig: context-map-meridiani}c, orange color unit) show the lowest
$c$ values ($ $$c_{CRISM}\sim0.50-0.60$, $\sigma\leq0.20$) and 
lower $c$ values than the equivalent unit in area 2 ($ $$c_{CRISM}\sim0.70-0.75$, 
$\sigma\leq0.20$). Compared to the in situ photometric results 
from Pancam measurements \citep{Johnson2006b}, higher $c$ values 
are observed than those estimated from Pancam for typical Outcrop units in Eagle crater and Tipuna (Figure 
\ref{fig: in-situ-parameters-meridiani}c). This difference can be explained by the difference of the 
spatial resolution and sampling strategy used for the different geological units.  Whereas CRISM 
pixels integrate over several geological units (basaltic granular materials, hematitic concretions, 
dust and sulfate-rich outcrops), Pancam data were extracted from individual portions of Outcrop blocks, 
typically without significant surface relief \citep{Johnson2014}. The MI images of the sulfate-rich outcrops 
(Figure \ref{fig: in-situ-observation-meridiani}b) showed evidence of heterogeneity in the 
sedimentary rocks which creates high scattering, consistent with a 
high $c$ value.

\subsubsection*{3.2.3.3. The surface macroscopic roughness parameter \label{sub:MER Opportunity theta}}

Figure \ref{fig: orbital-parameters-meridiani-1}d represents the
map of the surface macroscopic roughness parameter ($\bar{\theta}$)
values. We observe high
$\bar{\theta}$ values ($\sim$15-25$^\circ$), higher than those
estimated at Gusev plain. We note that: (i) the
region with a high extended area of bright-toned outcrops and accompanied
with mid-toned materials (Figure \ref{fig: context-map-meridiani}c,
orange color unit) and the region mainly composed
of mid-toned materials (Figure \ref{fig: context-map-meridiani}c,
yellow color unit) are associated with the highest $\bar{\theta}$
values ($\bar{\theta}_{CRISM}\simeq$ 15-25$^\circ$, $\sigma\leq$5$^\circ$
, Figure \ref{fig: orbital-parameters-meridiani-1}d), (ii) the
Botany Bay region, composed of dark-toned outcrops (Figure
\ref{fig: context-map-meridiani}c blue color unit) and the dark-toned
material mantle in the crater floor in area 2 (Figure \ref{fig: context-map-meridiani}c,
green color unit) show the lowest $\bar{\theta}$ values ($\bar{\theta}_{CRISM}\simeq$ 5$^\circ$, $\sigma\leq$5$^\circ$,
Figure \ref{fig: orbital-parameters-meridiani-1}d).

To examine the high $\bar{\theta}$ values, the mean slope is calculated
from HiRISE DTM, available in the area 2 at scale of 1 meter per pixel
(Figure \ref{fig: context-map-meridiani}c). By comparing the $\bar{\theta}$
parameter map to the mean slope, the high $\bar{\theta}$ values are
not correlated with a high mean slope and confirm again that the
macroscopic roughness parameter is more sensitive to the microscopic
topography (from the particle to a few mm). 

The macroscopic roughness values estimated by CRISM
are comparable to those estimated from Pancam measurements at 753 nm
\citep{Johnson2006b} (Figure \ref{fig: in-situ-parameters-meridiani}b).

\begin{figure}[H]
\begin{centering}
\includegraphics[scale=0.35]{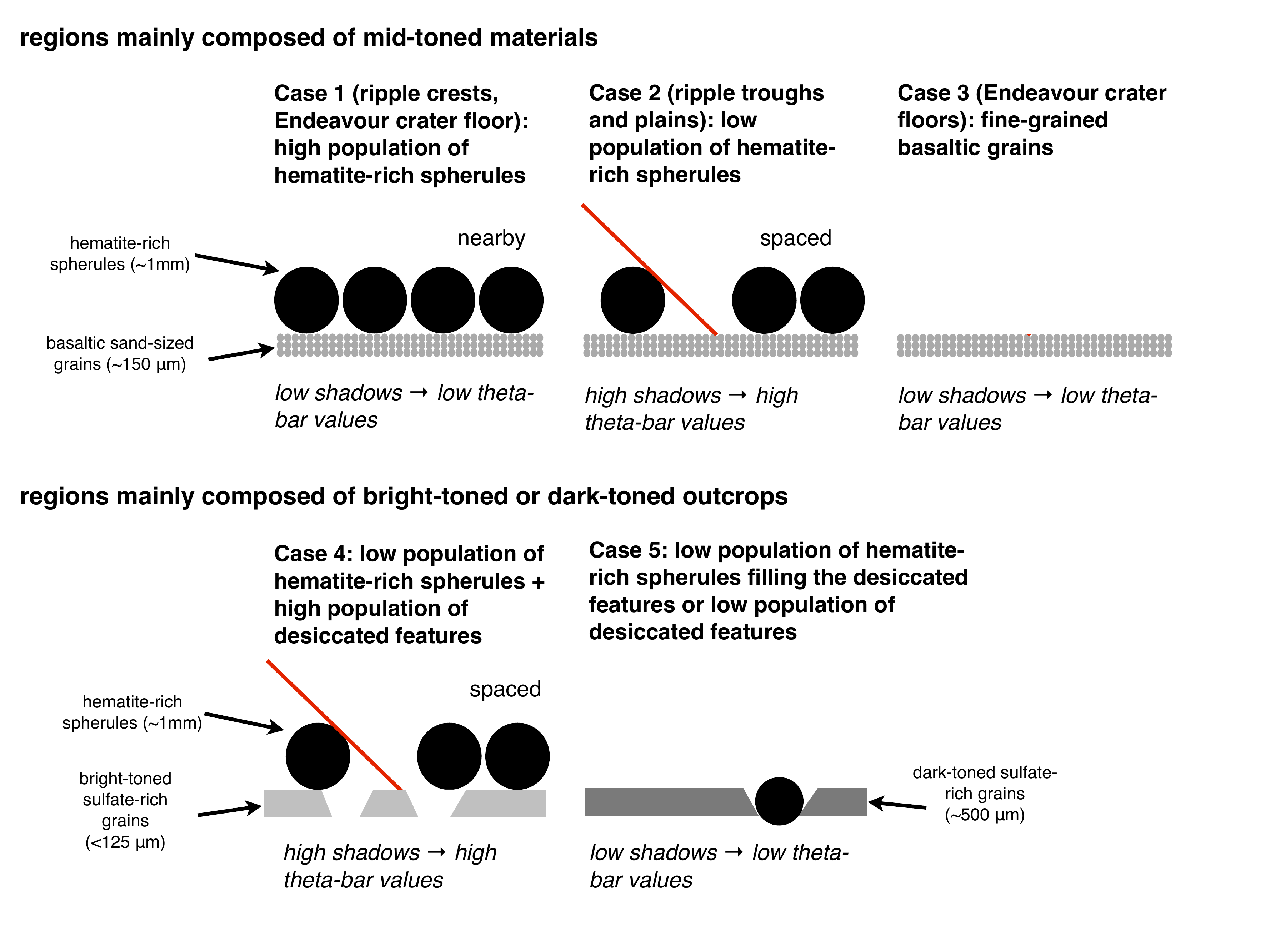}
\par\end{centering}

\caption{Schematic representation of the meaning of the macroscopic roughness values 
and the relation to the surface roughness in Meridiani Planum in regions dominated by 
dark-toned materials and bright-toned or dark-toned outcrops. \label{fig: roughness - meridiani}}

\end{figure}

(1) The regions mainly composed of mid-toned materials
(Figure \ref{fig: context-map-meridiani}c, yellow color unit) show
higher $\bar{\theta}$ values ($\bar{\theta}_{CRISM}\simeq$ 15-25$^\circ$) 
than in situ photometric results (Figure \ref{fig: in-situ-parameters-meridiani}b) for 
Spherule soil and Ripple soil units ($\bar{\theta}$ averages of $\sim$13$\pm$7 
and $\sim$10$\pm$1 $^\circ$, respectively). In situ observations showed that most of 
Meridiani soils (in plains and the ripple troughs which are the most sampled in a CRISM pixel) 
are composed of spaced millimeter-sized hematite spherules above a sand-sized basalt 
deposit (Figure \ref{fig: in-situ-observation-meridiani}b, top row), as represented in 
Figure \ref{fig: roughness - meridiani}, case 2. This particle distribution creates a high 
shadow hiding which explains the high $\bar{\theta}$ values of the CRISM and Pancam results. 
\citet{Johnson2013} showed in
their experimental studies on Mars analog materials that the addition
of spherules $<$2-3 mm in diameter with thin hematite rims, on an analog
of the spherule-bearing sulfate sedimentary rock much finer particles
(silt/sand size), caused the increase of the $\bar{\theta}$ values
(from 7$^\circ$ to 26$^\circ$). Those results are consistent with those observations.
Moreover, we can notice slightly higher $\bar{\theta}$ values for the Spherule soils unit 
than for the Ripple soils unit that can be explained by higher area density of spherules 
in the Ripple units (Figure \ref{fig: in-situ-observation-meridiani}b, top row) on average. 
Consequently, the high average CRISM $\bar{\theta}$ estimates, compared to those derived 
from Pancam measurements can be explained by a more extended sampling of the Spherule 
soils unit. Indeed, Spherule soils unit represents the most extended unit in the landscape, 
compared to the Ripple unit (Figure \ref{fig: in-situ-observation-meridiani}b, top row). 

(2) The regions with a high extended area of bright-toned
outcrops and accompanied with mid-toned materials (Figure \ref{fig: context-map-meridiani}c,
orange color unit) show consistent $\bar{\theta}$ values with the
in situ photometric results (Figure \ref{fig: in-situ-parameters-meridiani}b) 
for Outcrop units ($\bar{\theta}=$15-20$^\circ$).
Again, the presence of small population of hematitic concretions on
the sulfate-rich outcrops can create high shadow hidings which can explain
the high $\bar{\theta}$ values, as shown in Figure \ref{fig: roughness - meridiani}, case 4. 
Moreover, in situ observations showed that the sulfate-rich outcrops are highly textured 
(broad and weakly filled by hematite-rich spherules) which creates a high local slope. 
This high local slope can explain the high $\bar{\theta}$ values 
(Figure \ref{fig: roughness - meridiani}, case 4). Furthermore, high $\bar{\theta}$ values 
are noticed in the aeolian ripples with sulfate-rich outcrops in area 2, compared to those in 
the area 1 (Figure \ref{fig: context-map-meridiani}c, in orange color unit). This can be 
explained by higher exposure of sulfate-rich outcrops in area 2, due to the removal of superficial dust, also observed
in the npOx map (Figure \ref{fig: orbital-parameters-meridiani-1}b).
The TI values estimated from THEMIS measurements are higher in
the aeolian ripples with sulfate-rich outcrops in the area 2 (THEMIS TI values: 
$\sim$155-180 $J.m^{-2}.K^{-1}.s^{-1/2}$ \citep{Arvidson2011}) than those observed in the area
1 (THEMIS TI values: $\sim$140 -145 $J.m^{-2}.K^{-1}.s^{-1/2}$ \citep{Arvidson2011}), 
consistent with the $\bar{\theta}$ values. Those observations are consistent with our results.

(3) The region composed of dark-toned sulfate rich outcrop (Botany Bay) (Figure \ref{fig: context-map-meridiani}c, blue color unit) shows the lowest $\bar{\theta}$ values ($\bar{\theta}\sim$
5$^\circ$, $\sigma\leq$5$^\circ$, Figure \ref{fig: orbital-parameters-meridiani-1}d).
The $\bar{\theta}$ values can be explained by (i) small population of hematite-rich spherules atop the outcrops, and/or (ii) small population of desiccated features or narrower and/or (iii) the latter are filled by hematite-rich spherules which create a smooth area (Figure \ref{fig: roughness - meridiani}, case 5).

(4) The dark-toned material mantle (hematite-rich spherules and basaltic sand-sized grains) in the crater
floor (Figure \ref{fig: context-map-meridiani}c, green color unit
in area 2) shows the lowest $\bar{\theta}$ values ($\bar{\theta}\sim$
5$^\circ$, $\sigma\leq5$ $^\circ$, Figure \ref{fig: orbital-parameters-meridiani-1}d).
In situ observations are not available in this area, but we suggest that the low $\bar{\theta}$ 
values can be the result of: (i) the high density of hematitic concretions disposed nearby, if 
hematite-rich spherules cover the basaltic medium (Figure \ref{fig: roughness - meridiani}, case 1), or (ii) if hematite-rich spherules are absent, 
the presence of basaltic fine-grained materials creating a smooth surface (Figure \ref{fig: roughness - meridiani}, case 3).  


\section{Modeling \label{sec:Modeling}}

In the Section 3, photometric results were obtained in the MER landing site regions and discussed by 
comparing the in situ and orbital observations. However, several questions remain. 
Indeed, we observed from the in situ images that the surface is mainly composed of material mixtures. 
Consequently, it is important to know what kind of particles can control the overall photometric behaviors 
in a mixture and how we can relate the photometric results to the surface mixtures. 
Moreover, the information about the particle internal structure, derived from the single 
scattering albedo and phase function parameters, can be useful to constrain the geological processes (e.g., the presence 
or the absence of an internal structure in an igneous material can inform us about the 
crystallization mode). 

The radiative transfer model and analytic approaches are helpful to analyze the remote 
sensing measurements, to relate them in terms of surface material properties, and to reproduce 
realistic situations allowing us for a better understanding of the physical properties. 

To better understand the relation between the single scattering albedo parameter and the 
microscopic properties of the particles such as the composition (complex indices of refraction), 
the particle diameter (path length) and on the particle internal structure (density of 
internal scatterers), we used an analytic approach. In radiative transfer modeling, 
little is known about the range of internal scatterers coefficient ($s$) (see subsection 4.1 
and the Appendix A) values and they are usually set to 0 
for simplification \citep[e.g., ][]{Lucey1998}. Similarly, the relation between this coefficient and 
the internal structure of a particle is underconstrained \citep[e.g., ][]{Hapke2012}. Here, we estimated 
the internal scatterers coefficients of the observed compounds, at both MER landing sites
by using a combination of the CRISM single scattering albedo values (Section 3) and 
the in situ information in Section 4.1. 

Most commonly planetary surfaces are composed of a mixture of different materials. 
Consequently, the reflected radiation measurements from the CRISM instrument (spatial 
resolution: 200m/pixel) is a signal from different components of the surface. Moreover, 
little is known on what controls the overall photometric behavior of a mixture (e.g., spatial, 
intimate, stratified) made of different material properties. To understand the meaning of 
each photometric parameter estimated from CRISM observations and to evaluate which 
compound controls the overall photometric behavior, we performed realistic simulations by 
mimicking, as possible, the mixtures observed in the MER landing sites in Subsection 4.2. 

\subsection{Density of internal scatterers \label{sub:Internal scatterers} }

The theoretical formulation of Hapke (eq. 6.47-6.50,
\citet{Hapke2012a}, Appendix A) is used to relate the single scattering albedo
($\omega$) to the microscopic properties of particles, including the complex indices of 
refraction ($n$ and $k$) related to the particle composition, the path length ($D$) related 
to the particle size, the internal scattering coefficient ($s$) and the density of internal 
scatterers ($sD$) related to the particle internal structure. This formulation is used here 
for the calculation of the single scattering albedo value, knowing the particle properties.

\begin{figure}[H]
\begin{centering}
\includegraphics[scale=0.45]{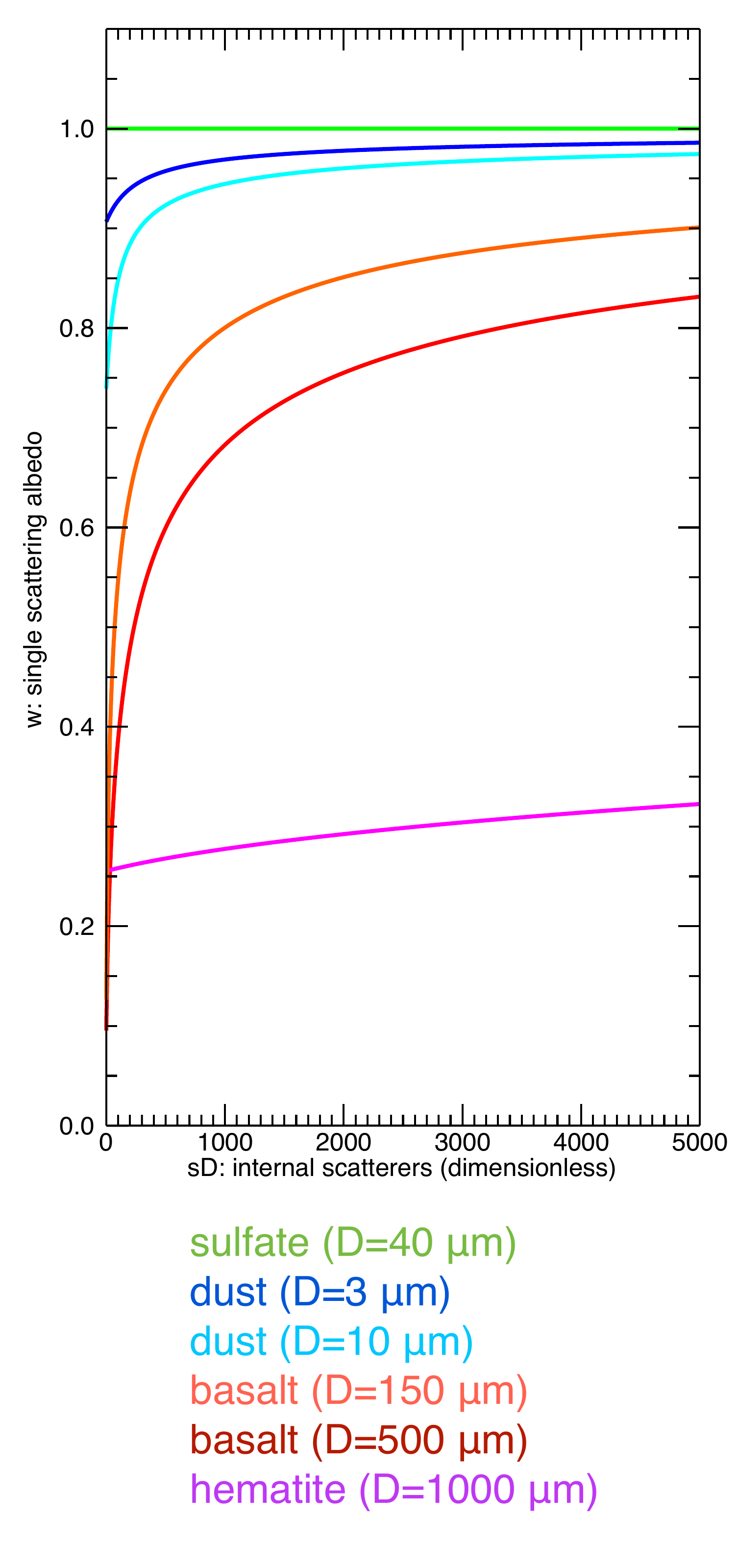}
\par\end{centering}

\caption{Graph of the calculated single scattering albedo ($\omega$) from the \citet{Hapke2012} 
formulation (eq. 6.47-6.50, Appendix A) as a function of the internal scattering coefficent ($sD$) for the typical
particle size of basalt (500 $\mu m$, in dark red color), of dust (3 $\mu m$,
in dark blue color and 10 $\mu m$, in light blue color), of hematite
(1000 $\mu m$, in purple color) and of sulfate (40 $\mu m$, in green
color). \label{fig:theoretical w}}
\end{figure}

For the case of MER-Spirit landing site, the $\omega$ values of basalt and nanophase ferric 
oxide dust are calculated from the optical constants, detailed in Table \ref{tab:Physical 
parameters - mixtures}. First, the $\omega$ value is calculated for typical dust, assuming 
isolated atmospheric particles with a diameter$ $ 3 $\mu m$ and no internal structure ($s=0$) 
\citep{lemmon2004}. This $\omega$ value is around 0.9 (Figure \ref{fig:theoretical w}), 
which is higher than the $\omega$ values estimated from CRISM observations in the dustier 
area in Figure \ref{fig: orbital-parameters-gusev}b ($\omega_{CRISM}\sim0.80$). 
The lower $\omega_{CRISM}$ value can be explained by aggregated dust with greater effective particle size. 
The $\omega$ value calculated for coarser dust with a diameter 10 $\mu m$ as a function 
of the density of internal scatterers ($sD$) is shown in Figure \ref{fig:theoretical w}. 
To have similar single scattering albedo values as the CRISM estimates ($\omega_{CRISM}\sim0.80$), 
internal scatterers must be included ($s=3.8\,\mu m^{-1}$ equivalent to $sD\sim40$). This 
result is consistent with the in situ observations showing that dust is aggregates
of unresolved subparticles \citep{herkenhoff2004a,Sullivan2008,Vaughan2010}.
Second, the $\omega$ value is calculated for typical
basalt ($\simeq$ 500 $\mu m$, Subsection \ref{sub:MER Spirit w},
\citep{herkenhoff2004a,Herkenhoff2006}) for different values of $s$
parameter. To have similar single scattering albedo values as the
CRISM estimates of the regions where the basaltic medium (less contaminated by dust) is visible in Figure \ref{fig: orbital-parameters-gusev}b ($\omega_{CRISM}\sim0.55$), $s=0.68\,\mu m^{-1}$ equivalent to $sD\sim$ 300, must be used, consistent with the high CRISM $c$ parameter values.

For the case of MER-Opportunity landing site, the $\omega$ values of hematite and sulfate are calculated from the optical constants and particle size, detailed in Table \ref{tab:Physical parameters - mixtures}.  Because the MI spatial resolution is limited to $\sim$100 $\mu m$ resolution, the component size of sulfate-rich outcrops is unknown. Consequently, we decided to refer to the \citet{Johnson2013}'s experimental works. They selected a hematite\textendash{}siderite spherule-bearing paleosol
as an analog to the spherule-bearing sulfate-rich sedimentary rocks
formed of a matrix soil of $<$ 45 $\mu m$ particles. For a typical
sulfate grain size ($\simeq$ 40 $\mu m$) and hematite grain
size ($\simeq$ 1000 $\mu m$) (because the hematite proportion in the spherules is unknown, we suppose that all the spherule is composed of hematite in first order), we note that the introduction of internal scatterers ($sD$) does not significantly change the $\omega$ value, compared to basalt and dust (Figure \ref{fig:theoretical w}). This
is due to the fact that sulfate is very bright, and that hematite has a very large grain size, and their physical properties take precedence over the density of internal scatterers.


\begin{sidewaystable}
\begin{centering}
{\scriptsize }%
\begin{tabular}{cc>{\centering}p{3cm}>{\centering}p{3cm}>{\centering}p{2cm}>{\centering}p{3cm}cc}
\hline 
{\scriptsize Site} & {\scriptsize Material} & {\scriptsize $n$} & {\scriptsize $k$} & {\scriptsize $s$ ($\mu m^{-1}$)} & {\scriptsize $D$ ($\mu m$)} & {\scriptsize $b$ } & {\scriptsize $c$ }\tabularnewline
\hline 
\multirow{2}{*}{{\scriptsize Spirit}} & {\scriptsize dust} & {\scriptsize 1.5 \citep{wolff2009}} & {\scriptsize 0.001 \citep{wolff2009}} & {\scriptsize 3.8 }{\scriptsize \par}

{\scriptsize (Subsection \ref{sub:Internal scatterers})} & {\scriptsize 10}{\scriptsize \par}

{\scriptsize \citep{lemmon2004}} & {\scriptsize 0.3} & {\scriptsize 0.6$^{1}$ \citep{Johnson2006a}}\tabularnewline
 & {\scriptsize basalt} & {\scriptsize 1.52 \citep{Pollack1973}} & {\scriptsize 0.0011 \citep{Pollack1973}} & {\scriptsize 0.68}{\scriptsize \par}

{\scriptsize (Subsection \ref{sub:Internal scatterers})} & {\scriptsize 500}{\scriptsize \par}

{\scriptsize \citep{Herkenhoff2006}} & {\scriptsize 0.3} & {\scriptsize 0.8 (Subsection \ref{sub:MER spirit parameters})}\tabularnewline
\multirow{3}{*}{{\scriptsize Opportunity}} & {\scriptsize hematite} & {\scriptsize 2.805 \citep{Sokolik1999}} & {\scriptsize 0.03 \citep{Sokolik1999}} & {\scriptsize 0}{\scriptsize \par}

{\scriptsize (Subsection \ref{sub:Internal scatterers})} & {\scriptsize 1000}{\scriptsize \par}

{\scriptsize \citep[e.g., ][]{Herkenhoff2008}} & {\scriptsize 0.3} & {\scriptsize 1.0 (Subsection \ref{sub:MER Opportunity parameters})}\tabularnewline
 & {\scriptsize basalt} & {\scriptsize 1.52 \citep{Pollack1973}} & {\scriptsize 0.0011 \citep{Pollack1973}} & {\scriptsize 0.68}{\scriptsize \par}

{\scriptsize (Subsection \ref{sub:Internal scatterers})} & {\scriptsize 150}{\scriptsize \par}

{\scriptsize \citep[e.g., ][]{Herkenhoff2008}} & {\scriptsize 0.3} & {\scriptsize 0.8$^{3}$ (Subsection \ref{sub:MER spirit parameters})}\tabularnewline
 & {\scriptsize sulfate} & {\scriptsize 1.5 \citep{Roush2007}} & {\scriptsize 10$^{-5}$ \citep{Roush2007}} & {\scriptsize 0}{\scriptsize \par}

{\scriptsize (Subsection \ref{sub:Internal scatterers})} & {\scriptsize 40\citep{Johnson2013}} & {\scriptsize 0.3} & {\scriptsize 0.2$^{2}$ \citep{Johnson2013}}\tabularnewline
\hline 
\end{tabular}
\par\end{centering}{\scriptsize \par}

{\scriptsize $n$: refractive index, $k$: absorption coefficient,
$s$: internal scattering coefficient, $D$: path length related to
the particle size, $b$: the asymmetry parameter from the two-term
Henyey Greenstein function, c: backscattering fraction from the two-term
Henyey Greenstein function}{\scriptsize \par}

{\scriptsize $^{1}$similar to the scattering properties of Bright
soil unit of the Landing site obtained from Pancam \citep{Johnson2006a},
corresponding to the most dustier area)}. {\scriptsize \par}

{\scriptsize $^{2}$based on \citet{Johnson2013}'s experimental photometric
results on an analog to the spherule-bearing sulfate sedimentary rocks
observed at MER-Opportunity landing site}{\scriptsize \par}

{\scriptsize $^{3}$same $b$ and $c$ values estimated for the basaltic granular medium
for the case of Gusev crater}{\scriptsize \par}

\caption{Physical parameters at 750 $nm$ used for the calculation of the $s$ parameter in Section 4.1 and for the mixture simulations in Section 4.2. The justifications are discussed in the text in Section 4.2. \label{tab:Physical parameters - mixtures}}
\end{sidewaystable}


\subsection{Mixtures \label{sub:Mixtures} }

In natural environments, planetary surfaces are composed of a
mixture of different materials (e.g., with different grain size, composition, internal structure). 
For instance, at the selected areas at the MER landing sites, several kind of mixtures 
are observed: dust deposits on basaltic unconsolidated materials observed at the MER-Spirit 
landing site and lags of hematite concretions on basaltic unconsolidated materials observed at the MER-Opportunity landing site. 

However, little is known about the influence of each compound of the mixture 
on the reflected radiation and more specially on the overall phase curve. By using a new 
radiative transfer model \citep{Pilorget2013b}, \citet{Pilorget2014} studied the evolution 
of the phase curve for various kinds of mixtures (spatial, intimate, and layered). The authors 
noticed that the phase curve evolution is controlled by the most abundant / brightest / highly 
anisotropic scattering grains within the mixture. The spatial and intimate mixtures showed 
similar trends in the phase curves when the grain properties varied. For the case of layered mixtures, 
the overall phase curves were generally very sensitive to the grain properties of the top layer.

In order to interpret our photometric results at the MER landing sites, we performed realistic 
simulations by attempting to mimic the observed mixtures. For that purpose, we used the 
same radiative transfer model as for the \citet{Pilorget2014}'s work \citep{Pilorget2013b} 
which simulates light scattering in compact granular media using a Monte-Carlo approach. 
The physical and compositional properties can be specified at the grain scale, allowing 
different kinds of mixtures (spatial, intimate and stratified). 
The radiative transfer is calculated by using a ray tracing approach between the
grains and probabilistic physical parameters, such as a single scattering
albedo and a phase function, at the grain scale. The single scattering
albedo is calculated using the \citet{Hapke1993,Hapke2012a} formulation (Appendix A) and, thus
is a function of the complex optical index of the material, the grain
size and the potential inclusion of internal scatterers. A two-lobe
Henyey-Greenstein phase function is also used in the model, consistent
with the method described in the previous sections. 

The photons' wavelength is 750 nm and their incidence angle is set at 45$^\circ$. A porosity
of 0.5 is assumed. The reflectance factor is computed in the -80,
+80$^\circ$ emergence range. These results are then inverted
by comparing and fitting the phase curves with the ones of homogeneous
samples with various single scattering albedos and phase functions.
Uncertainties are estimated to be within the $\pm$ 0.01 range for
the parameter $c$ and $\pm$ 0.02 for the parameter $\omega$. 

\begin{figure}[H]
\begin{centering}
\includegraphics[scale=0.35]{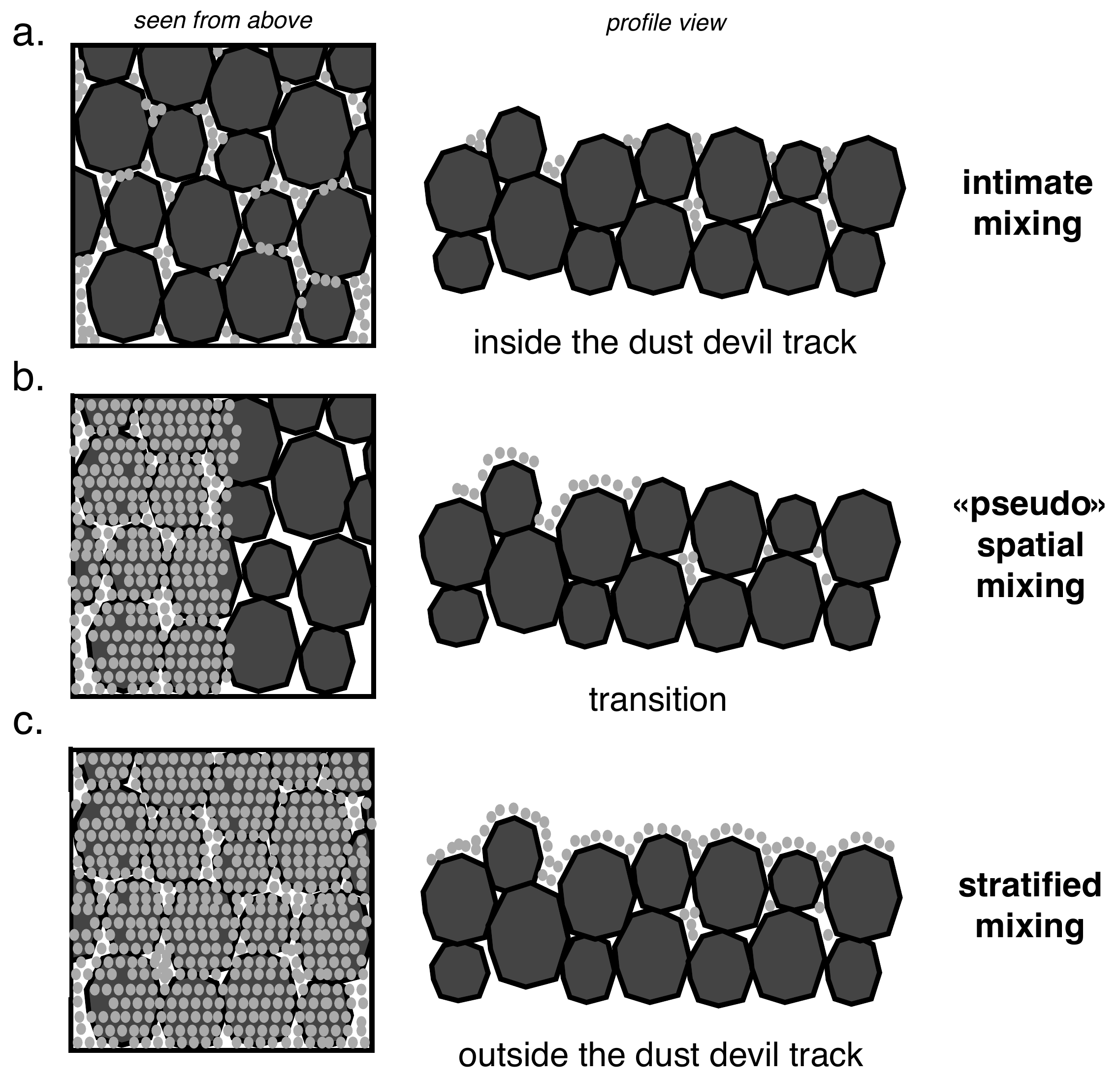}
\par\end{centering}

\caption{Schematic representation (seen from above and profile view) of different mixtures tested
with the numerical model for the case of MER-Spirit area (not at scale): (a) the
intimate mixture observed inside a dust devil track, (b) ``pseudo''
spatial mixture in dust devil track transition, (c) the stratified
mixture observed outside a dust devil track. The particle sizes used here are: 500 $\mu m$ 
for basaltic grains and 10 $\mu m$ for dust (assumed as a clumping of small dust).  \label{fig:mixture gusev scheme}}
\end{figure}

In the MER-Spirit landing site at Gusev Crater, dust may cover a
surface made of basalt grains or even fill the space pores between
them in regions affected by dust devils (Figures \ref{fig:mixture gusev scheme}).

\begin{figure}[H]
\begin{centering}
\includegraphics[scale=0.45]{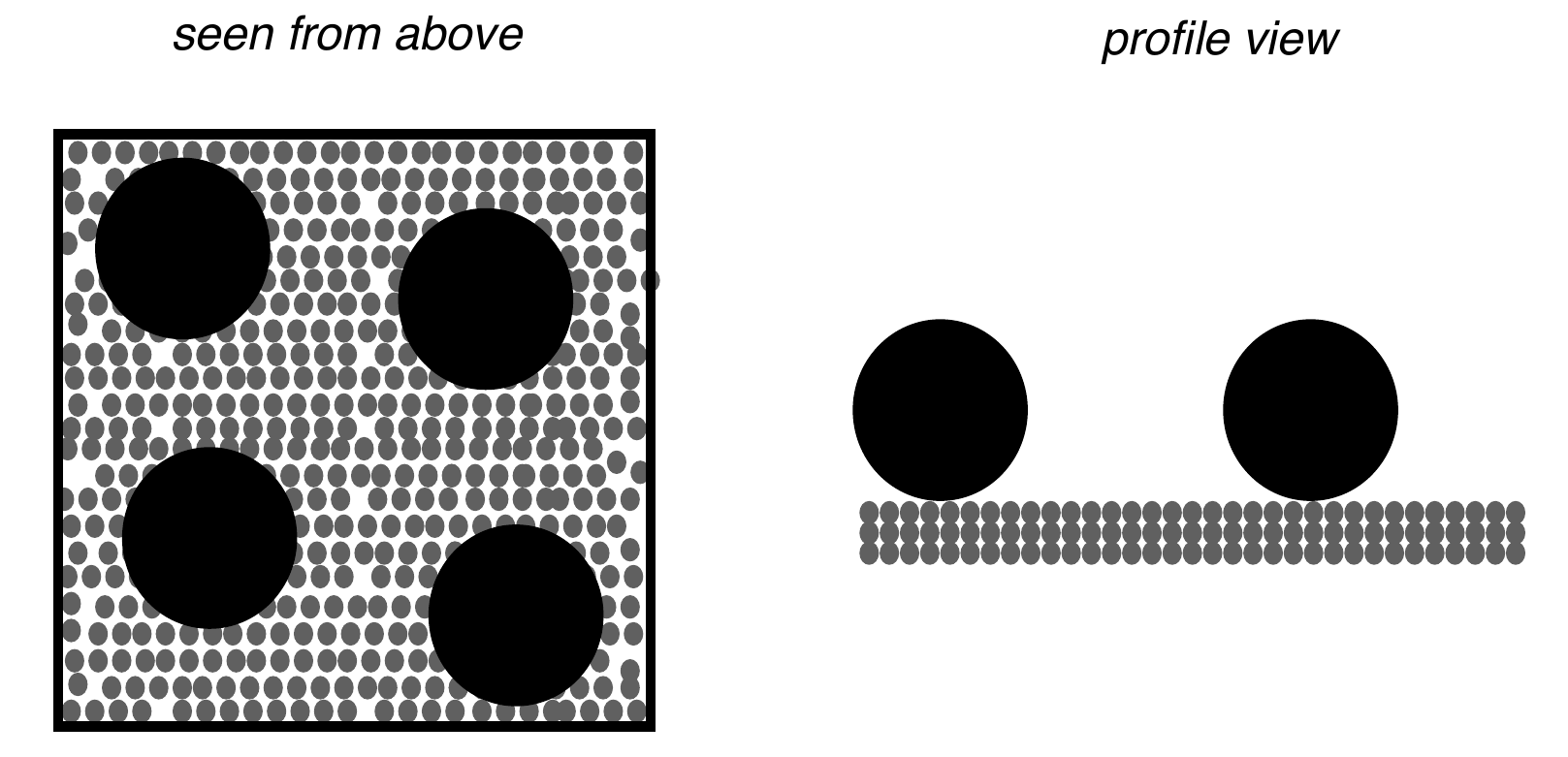}
\par\end{centering}

\caption{Schematic representation (seen from above and profile view) of mixtures tested with
the numerical model for the case of MER-Opportunity area (not at scale): hematitic
concretions (coarse black grains) on basalt or on sulfate layers (finer
grey grains).The particle sizes used here are: 1000 $\mu m$ for concretions, 
500 $\mu m$ for basaltic grains and 40 $\mu m$for sulfate grains. \label{fig:mixture meridiani scheme}}
\end{figure}

In the MER-Opportunity landing site at Meridiani Planum, hematitic
concretions may cover the surface composed of basaltic grains 
or composed of sulfate-rich outcrops (Figure \ref{fig:mixture meridiani scheme}).


For the simulations, several parameters must be known for each component of the mixture, 
but little is known about their complex indices of refraction, internal 
structure, shape and roughness. We decided to choose the following values:

(1) The composition, described by the complex indices of refraction ($n$ and $k$), can be 
estimated from analogous samples from laboratory measurements (e.g., for basaltic, hematite, 
sulfate materials) or from spaceborne observations (e.g., for dust). We assumed the hematite 
fraction in the concretions was 100\%. 

(2) The density of internal scatterers (noted $s$) is generally unknown for any samples 
\citep[e.g., ][]{Lucey1998}. This parameter was evaluated in Subsection 4.1. 

(3) The particle size (noted $D$) can be evaluated from in situ observations from the MI 
instruments (e.g., for basaltic and hematite materials) as long as the particle size is 
greater than the spatial resolution of the instrument. For instance, the dust and sulfate 
grain sizes are unknown. The dust in Gusev Crater was observed as a clumping of individual 
dust grains forming coarse aggregates in local areas along the rover traverse ($\leq$150 $\mu m$) 
\citep{Sullivan2008,Vaughan2010}. However, due to the spatial resolution limitation of the 
MI instrument, smaller aggregates may exist. We decided to simulate with a mean value of dust 
aggregates set at 10 $\mu m$. For the sulfate grains, we decided to choose the similar particle 
size used in the \citet{Johnson2013}'s experimental work on analogous samples (40 $\mu m$). 
In their study, the authors showed similar photometric results with those from in situ measurements. 

(4) The particle phase function of each compound is unknown. For the Gusev crater case, 
the parameters $b$ and $c$ of dust are similar to the scattering properties of Bright soil 
unit of the Landing site obtained from Pancam \citep{Johnson2006a}, corresponding to the most 
dusty area. For the basaltic grains, the phase function parameters $b$ and $c$ are the same 
as those estimated for the basaltic granular medium for the case of Gusev crater from CRISM 
data, corresponding to values less influenced by dust, compared to the in situ measurements 
where the media are usually covered by a thin dust layer \citep{Johnson2006a}. For the 
Meridiani Planum case, the parameters $b$ and $c$ of sulfate material are those estimated 
by \citet{Johnson2013}. The phase function parameters $b$ and $c$ for the basaltic sands are 
assumed to be same as those at the Gusev Crater plain. For the hematite-rich spherules, as 
discussed previously, the concretions may be composed of internal structures, such as an 
internal growth (concentric or radial) \citep{Glotch2006,Golden2008}, creating a high density 
of internal scatterers. The associated scattering behavior is therefore assumed broad and backscattered. 

All the cases are simulated using the input parameter values summarized in 
Table \ref{tab:Physical parameters - mixtures}.

\begin{figure}[H]
\begin{centering}
\includegraphics[scale=0.45]{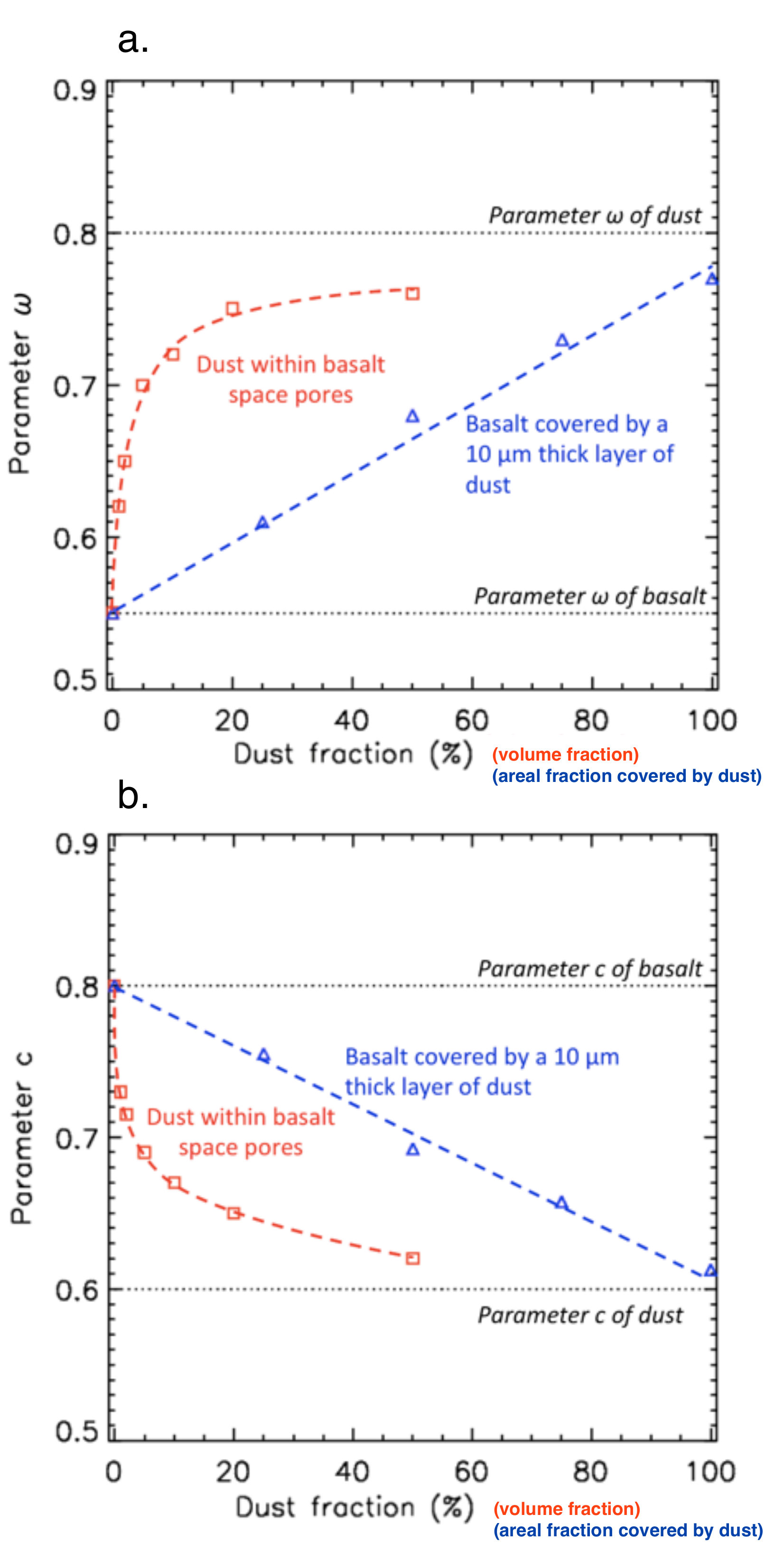}
\par\end{centering}

\caption{(Top) Evolution of the overall single scattering albedo (or parameter $\omega$) and 
(Bottom) evolution of the overall backscattering fraction (or parameter $c$) of the mixture 
when adding dust (assumed as a clumping of small dust particles) to a soil made of basaltic 
granular medium. Both intimate (in red) and layered cases (in blue) are simulated. The grain 
sizes, optical parameters and phase functions used here are summarized in Table \ref{tab:Physical parameters - mixtures}.
For the intimate case, dust grains are set between the basaltic grains (Figure \ref{fig:mixture gusev scheme}a). 
For the layered case, dust grains have a 10 microns size on top of the basaltic granular medium 
(Figures \ref{fig:mixture gusev scheme}b and c). Since the porosity is set at 0.5, 
there are therefore always some voids
between the dust grains. \label{fig:mixture gusev result}}
\end{figure}

Figure \ref{fig:mixture gusev result} represents the simulation for Gusev Crater showing 
the evolution of the single scattering albedo (Figure \ref{fig:mixture gusev result}a) and the backscattering fraction parameter (Figure \ref{fig:mixture gusev result}b)
(the parameter $b$ is constant), when dust is added to a soil composed of basaltic sands. 
Dust is modeled as 10 $\mu m$  agglomerates of smaller dust particles and each agglomerate is 
modeled as a discrete solid volume with the parameters $b$ and $c$ as measured at insitu \citep{Johnson2006a} 
($b=0.3$, $c=0.6$ and $s=3.8$, Table \ref{tab:Physical parameters - mixtures}). Both intimate (Figure \ref{fig:mixture 
gusev scheme}a) (in red in Figure \ref{fig:mixture gusev result}) mixture and partial blanket case (Figure \ref{fig:mixture 
gusev scheme}c) (in blue in Figure \ref{fig:mixture gusev result}) are tested. For the intimate case, dust agglomerates 
are set between the basaltic grains like in Figure \ref{fig:mixture gusev scheme}a. 
For the layered case, dust agglomerates are set 
as a layer of a unique agglomerate, representing a 10 $\mu m$ thickness dust layer on top of the basaltic 
granular medium. In the simulations, a contiguous areal fraction (from 0 to 100\%) of the basaltic 
granular medium is covered by the dust (Figures \ref{fig:mixture gusev scheme}b and c). 

Simulations show that when dust is situated within the space pores (intimate
case), only a small volume fraction is sufficient to hide the photometric
response of the basaltic granular medium (Figures \ref{fig:mixture gusev scheme}c, 
\ref{fig:mixture gusev result}). When covering the basaltic granular medium by dust grains, 
the evolution of parameters $\omega$ and $c$ tends to be linear as the areal fraction 
covered by dust increased, which is consistent with some kind of checkboard mixture. 

Modeling results also show that a 10 $\mu m$ thick layer of dust has a strong impact on the photometric 
response of the underlying basaltic medium (Figures \ref{fig:mixture gusev scheme}c, 
\ref{fig:mixture gusev result}). When dust covers 100\% of the basaltic granular medium, 
the influence of basaltic grains is small, but still exists, since voids are present between the dust grains. Such results, 
however, depend on both porosity and the geometric configuration of the grains. A higher porosity, 
as could be the case in agglomerates, will tend to increase the influence of the underlying substrate \citep{Pilorget2014}. 
With 5\% of dust in volume fraction, the parameter $\omega$ 
of the mixture is $\sim$0.7 and the parameter $c$ is $\sim$0.69. We obtained similar values 
for the layered case when 70\% of surface area is covered by a 10 $\mu m$ thick layer of dust  
(Figure \ref{fig:mixture gusev result}). 

These results show that in the Gusev crater area, the presence of
dust (layer or in the space pores) drives the mean scattering behavior
and can mask the underlying material and the variability of scattering
properties of materials. The dust devils are not sufficient to remove
all dust (layer or in the space pores) from the surface materials
and dust grains remain between grains. To assess the scattering behavior of the underlying
basaltic granular medium, the area must be affected by stronger aeolian processes such as 
storms, to remove sufficiently dust both on the surface and within the substrate. 


\begin{figure}[H]
\begin{centering}
\includegraphics[scale=0.4]{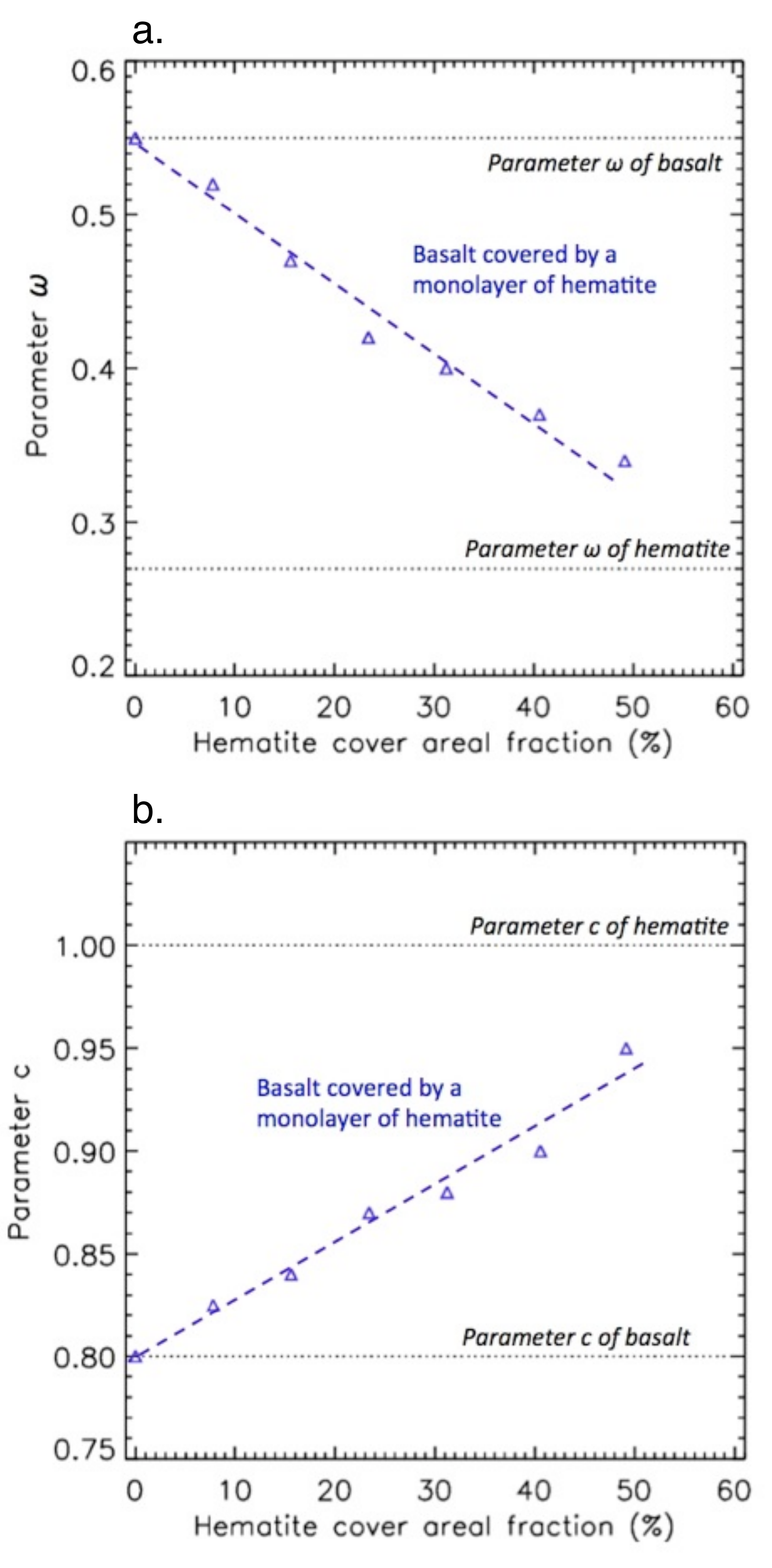}
\par\end{centering}

\caption{(Top) Evolution of the overall single scattering albedo ($\omega$)
and (Bottom) evolution of the overall backscattering fraction ($c$) of the mixture when 
adding hematite spheres (monolayer) on top of a soil made of basaltic sands. The grain sizes, 
optical parameters and phase functions used here are summarized in
Table \ref{tab:Physical parameters - mixtures}. Hematite spheres
are set randomly on top of the basalic grains (Figure \ref{fig:mixture meridiani scheme}). 
The areal fraction covered by hematite spheres represents the surface
occupied by the spheres when looking at nadir. \label{fig:mixture meridiani result 1}}
\end{figure}

\begin{figure}[H]
\begin{centering}
\includegraphics[scale=0.35]{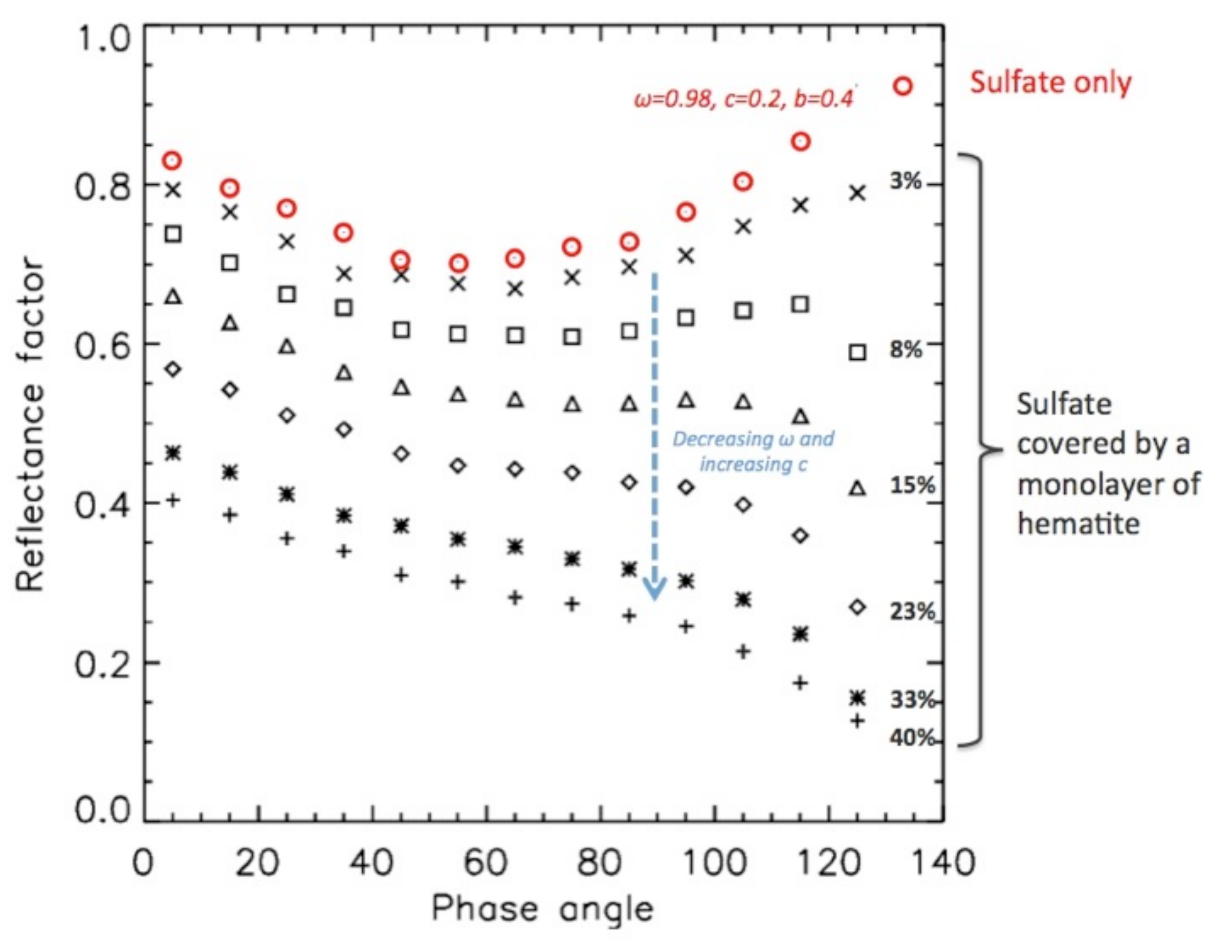}
\par\end{centering}

\caption{Evolution of the photometric curves when adding hematite spheres (monolayer) on 
top of a soil made of fine grained sulfates. The grain sizes,
optical parameters and phase functions used here are summarized in
Table \ref{tab:Physical parameters - mixtures}. Hematite spheres
are set randomly on top of the sulfate grains (Figure \ref{fig:mixture meridiani scheme}). 
The areal fraction covered by hematite spheres represents the surface
occupied by the spheres when looking at nadir. \label{fig:mixture meridiani result 2}}
\end{figure}

Figure \ref{fig:mixture meridiani result 1} represents the simulation for Meridiani Planum 
showing the simulation of the single scattering albedo (or parameter $\omega$) (Figure \ref{fig:mixture meridiani result 1}a) and evolution 
of the backscattering fraction (parameter $c$) (Figure \ref{fig:mixture meridiani result 1}b), when adding hematite spheres (unique layer of 
spherules) on top of a basaltic granular medium. Figure \ref{fig:mixture meridiani result 2} 
represents the evolution of the photometric curves, when adding hematite spheres on top of a 
soil made of fine grained sulfates. Hematite spheres are set randomly on top of the basaltic 
granular medium (Figure \ref{fig:mixture meridiani scheme}). Contrary to the case with a 
basaltic medium (Figure \ref{fig:mixture meridiani result 1}), geometric effects are very 
strong here, as demonstrated by \citet{Pilorget2014}. As the hematite sphere areal density increases, 
the reflectance drops at high emission angles (the photons that come from the underlying sulfates 
encounter the hematite spheres and are absorbed when escaping with high emergence angles). 
This effect is very strong when bright materials partially covering with a very absorbent 
material. No satisfactory fit could be obtained when trying to mimic this photometric 
behavior but several trends can be observed as hematite spheres are added onto the surface : (i) the general 
reflectance factor level tends to drop, resulting in a decrease of the single scattering albedo, 
(ii) the reflectance factor tends to drop more strongly for high phase angles, which tends to 
limit forward scattering and develop a backscattering behavior. Parameter $\omega$ drops 
from 0.98 (no hematite) to about 0.7 (40\% hematite cover).

Results show that the addition of hematitic spherules to the basaltic medium or to the 
sulfate medium decreases the surface brightness (Figures \ref{fig:mixture meridiani result 1}a 
and \ref{fig:mixture meridiani result 2}) and increases the $c$ values where the scattering 
became more backward (Figures \ref{fig:mixture meridiani result 1}b and \ref{fig:mixture meridiani result 2}). 
Similar results were observed by \citet{Johnson2013} where the additional of hematite-bearing 
spherules decreased the single scattering albedo
(from 0.97 to 0.84 at 750 nm) and increased the backscattering fraction (from
0.16 to 0.419 at 750 nm).

By comparing the CRISM estimates for in the regions dominated by hematitic concretions on 
basaltic sands ($\omega_{CRISM}=0.40-0.50$) to the simulations, results 
(Figure \ref{fig:mixture meridiani result 1}a) suggest that the basaltic layer on average may be 
covered by an areal fraction of hematite from 10 to 30\%. 

Moreover the modeling results suggest that regions where hematite-rich spherules 
cover the sulfate-rich deposits (Figure \ref{fig: context-map-meridiani}c, orange color unit), 
the spherules tend to control the overall photometric properties. By comparing the modeling results 
to the CRISM estimates, at least 40\% of hematite spherules on top of the sulfate medium are 
needed to have similar CRISM single scattering albedo ($\omega_{CRISM}=0.60-0.70$).


\section{Relations to the geological processes \label{sec:Geological processes}}

In the previous section, we underlined the main photometric results
and tried to provide a physical meaning in terms of particle size,
shape, internal structure and surface roughness. This section focuses
on the identification and the characterization of geological processes
from the surface photometric results.

\subsection{Gusev Crater}

\paragraph*{Volcanic resurfacing with the formation of the
primary basaltic crust}

The phase function parameters and the single scattering albedo values showed that the 
sand-sized basaltic granular medium (the dominating unit compared to the rocks in a CRISM 
pixel - 200 m/pxl) is composed of particles with moderate density internal scatterers. If 
the sand properties are similar to the properties of natural volcanic samples, then the 
sands may be characterized by internal structures such as micro-fractures, pores 
created by the escapement of gas during the magma crystallization, and/or crystals (microliths). 
The basaltic granular medium corresponds to the fractions of the basaltic rocks and crust which 
have been weathered. Due to the limitation of the spatial resolution of the MI instruments, 
the MI images cannot help us to understand the internal structure of the basaltic sands. 
However, the images of the basaltic rocks showed dark grey grains interpreted as olivine 
crystals (the biggest olivine crystals, larger than 100 $\mu m$ representing up to 20-30 volume \%) 
in an aphanitic groundmass \citep{McSween2006}. The phenocrysts and more specifically the crystals and glass particles 
in the groundmass may act as internal scatterers which can explain the phase function parameters 
 of the basaltic granular medium. This information provides key information about the magma 
 crystallization phase.

\paragraph*{Fragmentation of the volcanic crust by bolide impact}

The high macroscopic roughness values suggest the
presence of a high population of angular ejecta materials (millimeter-sized scale) which are associated with 
the numerous craters, consistent with the presence of spaced clasts
and rocks along the rover traverse \citep{Squyres2004a,grant2004,Grant2006,golombek2006a,arvidson2006a,ward2005}. 
This result reflects the fragmentation of the basaltic crust by local impacts and 
the emplacement of mm-sized ejecta materials in addition to the fragments resulting from 
typical physical weathering that comprise cm- and m-sized clasts and rocks.

\paragraph*{Aeolian processes with the reworking of the unconsolidated
materials}

The single scattering albedo values and the results presented in Section 4.1 indicate that dust seems to be
a clumping of individual dust particles forming coarse aggregates. Consequently, 
the subparticles can act like internal scatterers. This observation suggests that coarser particles than the individual dust particle with 3 $\mu m$ in diameter can be found. All these results are consistent with \citet{Sullivan2008,Vaughan2010}'s observations who identified
the presence of dust aggregates ($\leq$150 $\mu m$) in
local areas in the MER-Spirit traverse, on the rover magnets and on
solar panels. The aggregation of dust particles may be caused by electrostatic
forces \citep{Greeley1979}. The aggregates are very fragile, porous and 
weakly dense which make them easier to be entrained by wind than individual
dust particles or solid mafic grains of any large size
($\leq300\mu m$) \citep{Sullivan2008,Vaughan2010}. Then, these aggregates
disaggregate back into their $\leq3\,\mu m$ particles by wind \citep{Sullivan2008}
due to their fragility and the violence of the passing of dust devils
or strong winds contributing dust to the atmosphere. 
The removal of dust by aeolian processes (storms and dust devils)
exposes the lower-albedo basaltic media characterized by coarse grains. \citet{Thomas1984},
\citet{Edgett2000} and \citet{Greeley2005a} suggested that the removal
of a few micrometers thickness of dust from the surface could create
a reflectance difference as seen in the $\omega$ parameter map.
This observation is also strengthened by the modeling results in Subsection 4.2 
which showed that a dust cover can have a strong impact on the photometric 
response of the underlying basaltic medium.
The phase function parameter values show that the basaltic sands are
like rounded / spherical particles consistent with the in situ observations
\citep{greeley2006b,McGlynn2011}. As discussed before, the Gusev
basaltic crust was modified by bolide impacts. Consequently, the
generated granular materials would be angular, such as typical lunar
soils. The roundness of basaltic sands suggests that the materials
were transported in long distance by wind and modified by post-aeolian
processes \citep{McGlynn2011}. In the regions affected by dust devil
events, intermediate single scattering albedo values suggest that
dust remains on the surface materials. The MI observations showed
that dust infiltrates among the monolayer of basaltic sands which suggests
these sands are not currently experiencing saltation \citep{greeley2006b}.
The radiative transfer modeling of an intimate mixture between dust
and basaltic medium indicates that a small amount of dust can explain higher
$\omega$ values and lower $c$ values. Currently, only dust seems
to be mobilized by aeolian processes (e.g., dust devils).

\subsection{Meridiani Planum}

\paragraph*{Sulfate-rich deposits}

The ET2 unit is composed of sulfate-rich outcrops, identified in the
aeolian ripple troughs and near the Endeavour crater rim. From HiRISE,
the sulfate-rich outcrops in the ripple troughs are characterized
by bright-toned materials whereas the sulfate-rich outcrops near the
Endeavour crater are characterized by dark-toned materials. From photometric
parameters, we noticed differences in the scattering behaviors between
both etched terrains:

(1) The dark-toned etched terrains have lower $\omega$ than the bright-toned
terrains ($\omega=0.40-0.50$ for dark-toned etched terrains, $\omega=0.60-0.65$
for bright-toned etched terrains). This result can be explained by
the presence of coarser-grained particles in the dark-toned etched
terrains. 

(2) The dark-toned and bright-toned etched terrains have high $c$ ($c>0.75$ for dark-toned etched terrains, $c<0.75$ for bright-toned etched terrains). This suggests the presence of a heterogenous internal structures (even the first microns) created by mixtures of sulfates and very fine-grained siliciclastic materials, crystalline textures of subsequently precipitated cements and areas of recrystallization, centimeter-size vugs that record the early diagenetic growth and subsequent dissolution of crystals, and millimeter in size of spherules embedded in the outcrops \citep{herkenhoff2004b, Herkenhoff2008}. The internal heterogeneity may suggest a complex history of their formation, more especially during the precipitation and diagenesis phases. These phases may be composed of subsequent events of local recrystallization of cements and dissolution of the mineral grains and their evolution due to direct exposure and desiccation \citep{Squyres2004c}. Moreover, higher values of parameter $c$ are noticed for the dark-toned etched outcrops which may be explained by the presence of a greater heterogeneity. 

(3) The dark-toned etched terrains have lower $\bar{\theta}$ than
the bright-toned terrains ($\bar{\theta}=$5$^\circ$ for
dark-toned etched terrains, $\bar{\theta}=$15-25$^\circ$
for bright-toned etched terrains). This result can be explained by
the presence of flat sedimentary rocks, and/or less desiccated rocks
in the dark-toned regions.

The modeling results presented in Section 4.2 suggested that regions 
where hematite-rich spherules cover the bright sulfate-rich deposits 
tends to control the overall photometric properties, masking the photometric behavior of the underlying materials.

\paragraph*{Lag of hematitic concretions}

The Pm unit is characterized by a mixture of hematitic concretions
on basaltic sands. The single scattering albedo values and the phase
function parameter values underlined the presence of coarse spherical
hematitic concretions with moderate density of internal structure.


\section{Conclusion}

CRISM multi-angular observations allow for the characterization of the surface
scattering behavior, which depends on the composition but also the
material physical properties, such as the grain size and distribution, shape, internal
structure, and the surface roughness and porosity. First an atmospheric correction 
(compensating mineral aerosol effects) by the MARS-ReCO algorithm was used to estimate more accurately the
surface photometric curve taking into account the surface and mineral
aerosol scattering anisotropy. Then the surface photometric curve
was analyzed by inverting the Hapke photometric model depending on
six parameters: single scattering albedo, 2 phase function terms, macroscopic
roughness and 2 opposition effects terms, in a Bayesian
framework. Surface photometric maps were created
to observe the spatial variations of surface scattering properties
at the CRISM spatial resolution ($\sim$200m/pixel), as a function
of geological units. The orbital observations of the MER landing site 
regions in Gusev Crater and Meridiani Planum were used for
interpreting the estimated Hapke photometric parameters in terms of
physical properties, which can provide useful information about the geological
processes. The in situ observations were used as ground truth to validate
the interpretations. 

Variable scattering properties were observed within a CRISM observation
($\sim$5x10km) as a function of geological units suggesting variations
of surface physical properties. Such variations suggest that surfaces are controlled by
local geological processes rather than regional or global processes, like volcanic
resurfacing, fragmentation by impacts, aeolian processes, sulfates
deposition and diagenesis. 

Results are consistent with the in situ observations, thus 
validating the approach and the use of photometry for the estimates
of Martian surface physical properties. Some discrepancies resulted 
from a difference in the spatial scales. From the ground,
the in situ instruments can distinguish rocks and soils (centimeter
spatial scale), whereas CRISM observes extended areas (hectometric
spatial scale) composed of different components of the surface (rocks
and soils). 

Future works will focus on the determination of surface photometric
parameters of different geological terrains (sedimentary, volcanic
terrains, impact craters, etc), in order to identify variabilities
of scattering properties on Mars. 


\section*{Appendix A: Calculation of the single scattering albedo from \citet{Hapke2012a}'s formulation}

The single scattering albedo, $\omega$, is the ratio of scattered light at the particle
scale to extincted light and depends on the complex indices of refraction ($n$ and $k$)
related to the particle composition, the path length ($<D>$) related to the particle size, 
the internal scattering coefficient ($s$) and the density of internal scatterers ($sD$) related 
to the particle internal structure. The single scattering albedo varies from 0 (the light is 
totally absorbed) to 1 (the light is totally scattered) and it is defined as follows by \citet{Hapke2012}:

\begin{equation}
\omega=\frac{Q_{S}}{Q_{E}}=\frac{Q_{S}}{Q_{S}+Q_{A}}
\end{equation}

where $Q_{S}$ is the scattering efficiency of a particle and $Q_{E}$ the extinction efficiency of a particle

In the case of particles larger than the wavelength in a dense granular medium, the diffraction 
is ignored, so $Q_{E}=1$ \citep{Hapke2012a} and then: 

\begin{equation}
\omega=Q_{S}
\end{equation}

The scattering efficiency (nondiffractive) of large particles is the sum of two effiencies: 
(1) the total fraction of incident light that is reflected or emerges from the back surface of the slab 
(the surface facing the source) referred to as the backscattering efficiency, and (2) that emerging from the forward surface 
of the slab (the surface facing away from the source), referred to as the forward-scattering efficiency. \citet{Hapke2012a} 
(Eq. 6.20) provided the following equation for $Q_{S}$:

\begin{equation}
Q_{S}=S_{E}+\frac{(1-S_{E})\,(1-S_{I})}{1-S_{I}\,\Theta}\,\Theta
\end{equation}

\begin{itemize}
\item $S_{E}$ is the surface reflection coefficient for light that is externally incident 
on an irregular particle. \citet{Hapke2012a} (Eq. 6.49a) provided an analytic approximation 
to $S_{E}$ given by:

\begin{equation}
S_{E}=0.0587+0.8543\, R(0)+0.0870\, R(0)^{2}
\end{equation}

with $R(0)$, the normal specular reflection coefficient \citep{Hapke2012a} 
(Eq. 6.49a): $R(0)=\frac{(n-1)^{2}+k^{2}}{(n+1)^{2}+k^{2}}$

\end{itemize}

\begin{itemize}
\item $S_{I}$ is the surface reflection coefficient for light that is internally incident 
on a particle. An approximation to $S_{I}$ is given by \citet{Hapke2012a} (Eq. 6.23) as follows:

\begin{equation}
S_{I}=1-\frac{1}{n^{2}}\,(0.9413-0.8543\, R(0)-0.08710\, R(0)^{2})
\end{equation}

\end{itemize}

\begin{itemize}
\item $\Theta$ is the internal-transmission factor and the expression is \citep{Hapke2012a} (Eq. 6.30):

\begin{equation}
\Theta=\frac{r_{i}+exp(-\sqrt{(\alpha\times(\alpha+s)}\,<D>)}{1.+r_{i}\, exp(-\sqrt{(\alpha\times(\alpha+s)}\,<D>)}
\end{equation}

with $r_{I}$, the internal diffusive reflectance \citep{Hapke2012a} (Eq. 6.48b): $r_{I}=\frac{1-\sqrt{\alpha/(\alpha+s)}}{1+\sqrt{\alpha/(\alpha+s)}}$

with $\alpha$, the absorption coefficient: $\alpha=\frac{4\pi k}{\lambda}$

\end{itemize}

\begin{itemize}
\item $s$ is the internal scattering coefficient, expressed in length$^{-1}$. If $s=0$, that means that there are no internal scatterers inside the particle. The $s$ parameter describes the light attenuation by scattering caused by the presence of internal scatterers \citep{Hapke2012a}. 
\end{itemize}

\begin{itemize}
\item $<D>$ is the mean ray path length through particle in the case where there are no internal scatterers. If the particle is spherical then $<D>=0.9\times D$ where $D$ is the mean diameter of the particle. If the particle is irregular, $<D>$ could be quite different from $D$ (in general will be smaller) \citep{Hapke2012a}.
\end{itemize}


\paragraph{Acknowledgement} This work was supported by the French Space
Agency CNES (Centre National d'Etudes Spatiales) and PNP (Programme
National de Plan\'etologie) from INSU (Institut National des Sciences
de l'Univers). The authors would like to thank Michael Wolff for making
his aerosol optical thickness values available for this study, the CRISM team for providing the data. 
Finally, we would like to gratefully thank Jeffrey Johnson and the anonymous reviewer for 
their constructive comments which substantially improved the manuscript. 




\end{document}